\documentclass[aps,prd,twocolumn,groupedaddress,nofootinbib,amssymb,eqsecnum]{revtex4}

\usepackage{graphicx}
\usepackage{amsmath,amssymb,mathtools}

\DeclareMathOperator{\csch}{csch}

\usepackage{dcolumn}
\usepackage{hyperref}
\usepackage[pdftex]{color}
\usepackage[tight]{units}

\newcommand{\Order}{\mathcal{O}}

\newcommand{\OK}{\Omega_\mathrm{K}}
\newcommand{\OL}{\Omega_\Lambda}
\newcommand{\ODE}{\Omega_{\mathrm{QE}}}
\newcommand{\OQE}{\Omega_{\mathrm{QE}}}
\newcommand{\Or}{\Omega_\mathrm{r}}
\newcommand{\Ob}{\Omega_\mathrm{b}}
\newcommand{\Oc}{\Omega_\mathrm{c}}
\newcommand{\Om}{\Omega_\mathrm{m}}

\newcommand{\dQE}{\delta_{\mathrm{QE}}}
\newcommand{\wQE}{w_{\mathrm{QE}}}
\newcommand{\woQE}{w_{\mathrm{QE}}^0}

\newcommand{\dDE}{\delta_{\mathrm{DE}}}
\newcommand{\wDE}{w_{\mathrm{DE}}}
\newcommand{\woDE}{w_{\mathrm{DE}}^0}
\newcommand{\waDE}{w_{\mathrm{DE}}^a}

\newcommand{\xc}{\xi_\chi}
\newcommand{\xh}{\xi_h}

\newcommand{\ns}{n_s}
\newcommand{\as}{\alpha_s}
\newcommand{\Ninf}{N_{\mathrm{inf}}}

\newcommand{\cosmomc}{\textsc{CosmoMC}}
\newcommand{\camb}{\textsc{CAMB}}
\newcommand{\getdist}{\textsc{GetDist}}

\begin{document}

\title{Cosmological Constraints on Higgs-Dilaton Inflation}

\author{Manuel Trashorras}
\email{manuel.trashorras@csic.es}

\author{Savvas Nesseris}
\email{savvas.nesseris@csic.es}

\author{Juan Garc\'ia-Bellido}
\email{juan.garciabellido@uam.es}

\affiliation{Instituto de F\'isica Te\'orica UAM-CSIC, Universidad Auton\'oma de Madrid, Cantoblanco, 28049 Madrid, Spain}


\begin{abstract}

We test the viability of the Higgs-dilaton model (HDM) compared to the evolving dark energy ($w_0 w_a$CDM) model, in which the cosmological constant model $\Lambda$CDM is also nested, by using the latest cosmological data that includes the cosmic microwave background temperature, polarization and lensing data from the \textit{Planck} satellite (2015 data release), the BICEP and Keck Array experiments, the Type Ia supernovae from the JLA catalog, the baryon acoustic oscillations from CMASS, LOWZ and 6dF, the weak lensing data from the CFHTLenS survey and the matter power Spectrum measurements from the SDSS (data release 7). We find that the values of all cosmological parameters allowed by the Higgs-dilaton inflation model are well within the \textit{Planck} satellite (2015 data release) constraints. In particular, we have that $w_0 = -1.0001^{+0.0072}_{-0.0074}$, $w_a = 0.00^{+0.15}_{-0.16}$, $\ns = 0.9693^{+0.0083}_{-0.0082}$, $\as = -0.001^{+0.013}_{-0.014}$ and $r_{0.05} = 0.0025^{+0.0017}_{-0.0016}$ (95.5\%C.L.). We also place new stringent constraints on the couplings of the Higgs-dilaton model and we find that $\xc < 0.00328$ and $\xh / \sqrt{\lambda} = 59200^{+30000}_{-20000}$ (95.5\%C.L.). Furthermore, we report that the HDM is at a slightly better footing than the $w_0 w_a$CDM model, as they both have practically the same chi-square, i.e. $\Delta \chi^2 = \chi^2_{w_0 w_a\mathrm{CDM}}-\chi^2_{\mathrm{HDM}}=0.18$, with the HDM model having two fewer parameters. Finally Bayesian evidence favors equally the two models, with the HDM being preferred by the AIC and DIC information criteria.

\end{abstract}

\maketitle


\section{Introduction}
\label{sec:Introduction}

Often it has been successful in physics to try to relate apparently different phenomena in a common, unifying framework. Along these lines, the Higgs-dilaton model (HDM) was proposed in Refs. \cite{Shaposhnikov:2008xb,GarciaBellido:2011de,GarciaBellido:2012zu}. The model has its roots in a Higgs-inflation paradigm, see e.g. \cite{Salopek:1988qh,Bezrukov:2007ep}, and it is based on a nonminimal extension of the Standard Model (SM) to unimodular gravity (UG), a restricted version of general relativity (GR) in which the metric determinant is fixed to one, ${|g| = 1}$, and its main motivation is to explain the source of inflation while also relating the different scales of physics -- the Planck mass $M_P$, the Higg's vacuum expectation value $v$, and a possible cosmological constant $\Lambda$ -- that remain when nonminimally coupling the SM to GR/UG.

The HDM contains two scalar fields: the SM Higgs $h$ and a new dilaton $\chi$, both nonminimally coupled to gravity. It is by construction scale invariant (SI) at the classical level, and all the SM scales originate from the spontaneous breakdown of SI. It can explain inflation as a consequence of the slow roll of the $h$ and $\chi$ fields. Also, it provides a source of the dark-energy-driven expansion of the Universe (see Ref.~\cite{Bezrukov:2007ep}) once the fields reach their approximate ground state and start to roll down the potential valley. Therefore, the HDM bridges these two paradigms and uniquely provides an explanation of the current dark-energy-led accelerated expansion of the Universe from an inflationary paradigm.

The connection between the two eras allows us to relate inflation observables (namely, the primordial scalar power spectrum index $\ns$ and its running $\as$) to those observables associated with dark-energy (the dark energy equation of state parameter $\woDE$ and its evolution $\waDE$), as they all depend on the HDM parameters (the coupling constant of the Higgs, $\xh$, and the dilaton, $\xc$, to gravity, and Higgs' $\lambda$). Even though the cosmological constant model is very successful in describing the evolution of the Universe \cite{Ade:2015xua} and has passed many parametric \cite{Nesseris:2010ep} and nonparametric tests \cite{Nesseris:2012tt}, it still faces several apparently serious theoretical issues \cite{Bull:2015stt}, especially regarding to the predictivity and testability of the inflationary paradigm.

The aim of the present paper is to test the viability of the Higgs-dilaton model by comparing it to the cosmological constant ($\Lambda$CDM) and evolving dark energy ($w_0 w_a$CDM) model, by using the latest cosmological data that includes the cosmic microwave background (CMB) temperature, polarization and lensing data from the \textit{Planck} satellite (2015 data release), the BICEP and Keck Array experiments, the Type Ia supernovae from the JLA catalog, the BAO from CMASS, LOWZ and 6dF, the weak lensing data from the CFHTLenS survey and the matter power spectrum measurements from the SDSS (data release 7). To do this, we make the necessary modifications to the \cosmomc~suite in order to put constraints on the HDM parameters. By checking whether the model is compatible with present observations we will also provide insight on which future probes may be able to discern the model from the standard cosmological constant model $\Lambda$CDM and the evolving dark energy ($w_0 w_a$CDM) model in which is nested.

\subsection{Nonminimal SM coupling to general relativity}
\label{subsec:Non-minimal_SM_coupling_to_General_Relativity}

The simplest models of inflation are those sourced by a single scalar field and one may be tempted to identify such a field with the only scalar particle yet found, the Higgs boson. Such models, however, are not compatible with electroweak experiments and must be discarded, see Ref.~\cite{Nakamura:2010zzi}. The next simplest possibility is to introduce a new, massless scalar degree of freedom, hereafter called the dilaton. Neglecting the SM contributions, the scale-invariant extension for the SM plus GR including the dilaton is \cite{Shaposhnikov:2008xb}\footnote{Hereafter we follow convention $\eta_{\mu \nu} = \mathrm{diag} (-1,1,1,1)$ and $R^\alpha_{\beta \gamma \delta} = \partial_\gamma \Gamma^\alpha_{\beta \delta} + \Gamma^\alpha_{\lambda \gamma} \Gamma^\lambda_{\beta \delta} - (\gamma \leftrightarrow \delta$).}
	\begin{equation}
	\label{eq:Lagrangian_SI}
	\begin{split}
	\frac{\mathcal{L}_{SI}}{\sqrt{-g}} = 	& \frac{1}{2} \left( \xc \chi^2 + \xh h^2 \right) R
														\\
														& -\frac{1}{2} \partial^\mu h \partial_\mu h -\frac{1}{2} \partial^\mu \chi \partial_\mu \chi - V(h,\chi)
														\;,
	\end{split}
	\end{equation}
where the scalar potential is
	\begin{equation}
	\label{eq:Potential_SI}
	V(h,\chi) = \lambda \left( \frac{1}{2} h^2 - \frac{\alpha}{2 \lambda} \chi^2 \right)^2 + \beta \chi^4
	\;
	\end{equation}
and $h$ is the Higgs field in the unitary gauge. For the term multiplying the Ricci curvature in Eq.~\eqref{eq:Lagrangian_SI} to be positive, the couplings must also be $\xc, \xh \geq 0$. Furthermore, note that we have neglected a $R^2$ term in the lagrangian of Eq.~\eqref{eq:Lagrangian_SI} as it is equivalent to a new scalaron field under metric redefinitions. The scalaron field is degenerate with the Higgs nonminimal coupling, except for its gravitational interactions to the rest of matter.

The classical ground states are
	\begin{equation}
	\label{eq:Ground_States_SI}
	h_0^2 = \frac{\alpha}{\lambda} \chi_0^2 + \frac{\xh}{\lambda} R, \qquad R = \frac{4 \beta \lambda}{\lambda \xc + \alpha\xh} \chi_0^2,
	\;
	\end{equation}
which simplify greatly if $\beta = 0$, a parameter that determines whether spacetime is flat ($\beta = 0$), de Sitter ($\beta > 0$), or anti-de Sitter ($\beta < 0$) for a particular scalar curvature $R$ whose sign is controlled by $\beta$.

Note that the solutions for with $\chi_0 \neq 0$ spontaneously break scale invariance, and so all induced scales are proportional to $\chi_0$. In particular, the three SM induced scales are (see Ref.~\cite{GarciaBellido:2011de})
	\begin{align}
	\label{eq:Planck_Mass}
	M_P^2 = 	& \left( \xc + \xh \frac{\alpha}{\lambda} + \frac{4 \beta \xh^2}{\lambda \xc + \alpha \xh} \right) \chi_0^2
					\\
	\label{eq:Cosmological_Constant}
	\Lambda =	& \frac{\beta M_P^4}{(\xc + \alpha \xh / \lambda)^2 + 4 \beta \xh^2/ \lambda}
					\\
	\label{eq:Higgs_Mass}
	m^2_h = 	& 2 \alpha M_P^2 \frac{( 1 + 6 \xc ) + \alpha ( 1 + 6 \xh ) / \lambda}{\xc ( 1 + 6 \xc ) + \xh \alpha ( 1 + 6 \xh ) / \lambda}
					\;.
	\end{align}
	
One may worry how the introduction of a new scalar degree of freedom would alter the SM phenomenology, but, as shown in Ref.~\cite{Blas:2011ac}, the dilaton completely decouples from all Standard Model fields except for the Higgs, to which it is coupled derivatively, but Planck-suppressed by term of order $m_h^2 / M_P^2$. This means the dilaton does not alter the low energy SM phenomenology, thus making the HDM a viable effective field-theory extension of the Standard Model and general relativity.

\subsection{Non-minimal SM coupling to unimodular gravity}
\label{subsec:Non-minimal_SM_coupling_to_Unimodular_Gravity}

Now we replace general relativity by unimodular gravity, see Ref.~\cite{Weinberg:1988cp}, in which one reduces the independent components of the metric $g_{\mu\nu}$ by one, imposing the constrain of unit metric determinant $g \equiv \det(g_{\mu \nu}) = 1$, hence the name. This is not a strong condition, however, as it still allows the theory to describe all possible geometries. In this case the previous Lagrangian of Eq.~\eqref{eq:Lagrangian_SI} becomes\footnote{A hat on a quantity indicates that it depends on the unimodular metric $\hat g_{\mu \nu}$, satisfying the Unimodular Condition $\det \hat g_{\mu \nu} = -1$.}
	\begin{equation}
	\label{eq:Lagrangian_UG}
	\begin{split}
	\frac{{\cal L}_{SI + UG}}{\sqrt{-\hat{g}}} = 	& \frac{1}{2} \left( \xc \chi^2 + \xh h^2 \right) \hat{R} + {\cal L}_{\rm{SM}[\lambda \rightarrow 0]}
																\\
																& -\frac{1}{2} \hat{g}^{\mu\nu} \partial_\mu \chi \partial_\nu \chi - V(h,\chi) - \Lambda_0
																\;,
	\end{split}
	\end{equation}
where the potential $V(h,\chi)$ is still given by Eq.~\eqref{eq:Potential_SI}, and there appears a new $\Lambda_0$ term that explicitly breaks scale-invariance. At his point, we should note that adding a bare Higgs mass term $m^2h^2$ to Eq.~(\ref{eq:Lagrangian_UG}) is unnecessary, as it is already present through the $\alpha$ coupling to the dilation, whose SSB vev gives mass to the Higgs. The $\Lambda_0$ term is completely different, since it arises as a Lagrange multiplier related to the condition of Unimodular gravity and cannot be renormalized, thus alleviating the ``technical" cosmological constant problem \cite{Shaposhnikov:2008xb}.

Moving from the Jordan to the Einstein frame\footnote{A tilde on a quantity indicates that it is expressed on the Einstein frame, where $\tilde{g}_{\mu \nu} = \Omega^2 \hat{g}_{\mu \nu}$ and $\Omega^2 = M_P^{-2} f(\boldsymbol{\phi})$, $\boldsymbol{\phi} = (\phi^1,\phi^2) = (h,\chi)$.}, that is, making the metric transformation
	\begin{equation}
	\label{eq:Einstein_Jordan_Frames}
	\tilde{g}_{\mu \nu} = M_P^{-2} (\xc \chi^2 + \xh h^2) \hat{g}_{\mu \nu}
	\;,
	\end{equation}
causes the Lagrangian of Eq.~\eqref{eq:Lagrangian_UG} to transform into
	\begin{equation}
	\label{eq:Lagrangian_EF}
	\frac{{\cal L}}{\sqrt{-\tilde{g}}} = M_P^2 \frac{\tilde{R}}{2} + \tilde{{\cal L}}_{\rm{SM}[\lambda \rightarrow 0]} - \frac{1}{2} \tilde{K} - \tilde{U}(h,\chi)
	\;,
	\end{equation}
where $\tilde{K}$ is a noncanonical kinetic term,
	\begin{equation}
	\label{eq:Kinnetic_EF}
	\tilde{K} = \gamma_{a b} \tilde{g}^{\mu \nu} \partial_\mu \phi^a \partial_\nu \phi^b
	\;,
	\end{equation}
$\gamma_{a b}$ is a generally nondiagonal, noncanonical spatial metric,
	\begin{equation}
	\label{eq:Kinnetic_Gamma_EF}
	\gamma_{a b} = \frac{1}{\Omega^2} \left( \frac{3}{2} M_P^2 \frac{\partial_a \Omega^2 \partial_b \Omega^2}{\Omega^2}\right)
	\;,
	\end{equation}
and $\tilde{U}(h,\chi)$ is the scalar potential,
	\begin{equation}
	\label{eq:Potential_EF_beta_No_Cero}
	\begin{split}
	\tilde{U}(h,\chi) =	& \frac{M_P^4}{\left( \xc \chi^2 + \xh h^2 \right)^2}
								\\
								& \times\left( \frac{\lambda}{4} \left( h^2 - \frac{\alpha}{\lambda} \chi^2 \right)^2 + \beta \chi^4 + \Lambda_0 \right)
								\;.
	\end{split}
	\end{equation}

From here, it is straightforward to find the classical ground states for $\alpha$, $\lambda$, $\xc$, $\xh>0$. We do so first by finding them for the case where there is no cosmological constant $\Lambda_0$, which in any case is very small, and second by observing in them the effect of a nonzero cosmological constant.

If $\Lambda_0 = 0$, then the theory Eq.~\eqref{eq:Lagrangian_UG} reduces to Eq.~\eqref{eq:Lagrangian_SI}, and the potential is minimal along the two valleys
	\begin{equation}
	\label{eq:Ground_States_EF}
	h_0^2 = \left( \frac{\alpha}{\lambda} + \frac{4 \beta \xh}{\lambda \xc + \alpha\xh} \right) \chi_0^2
	\;.
	\end{equation}
	
The effect of a nonzero $\Lambda_0$, is to give the valleys a tilt, which breaks the degeneracy of the classical ground states, no longer flat. Two cases are then possible:
	\begin{itemize}
	\item For $\Lambda_0 < 0$ the potential valleys are tilted towards the origin. The true classical ground state for this case is a trivial nondegenerate one with $\chi = h = 0$.
	\item For $\Lambda_0 > 0$ the potential valleys are tilted away from the origin. There is an asymptotic classical ground state, given by Eq.~\eqref{eq:Ground_States_EF}, as the scalar fields go to infinity $\chi_0, h_0 \rightarrow \infty$. Depending on the sign of $\beta$, this asymptotic solution corresponds to a Minkowski, de Sitter (dS) or anti-de Sitter (AdS)  spacetime (see Eq.~\eqref{eq:Cosmological_Constant}).
	\end{itemize}
Therefore, $\Lambda_0 > 0$ does not play the role of a cosmological constant but rather gives rise to a run-away potential for the scalar fields. Moreover, if $\beta = 0$, dark energy does not contain a pure-constant contribution and is entirely generated by the term proportional to $\Lambda_0$. This latter condition on $\beta$ is assumed for the rest of the paper, and leads to the following scalar field potential Eq.~\eqref{eq:Potential_EF_beta_No_Cero} to take the simple form
	\begin{equation}
	\label{eq:Potential_EF_beta_Si_Cero}
	\begin{split}
	\tilde U(h,\chi) = 	& \frac{M_P^4}{\left( \xc \chi^2 + \xh h^2 \right)^2}
							\\
							& \times \left( \frac{\lambda}{4} \left( h^2 - \frac{\alpha}{\lambda} \chi^2 \right)^2 + \Lambda_0 \right)
							\;,
	\end{split}
	\end{equation}
and the potential valleys reduce to
	\begin{equation}
	\label{eq:Ground_States_EF_beta_Si_Cero}
	h_0^2 = \frac{\alpha}{\lambda} \chi_0^2
	\;.
	\end{equation}

\subsection{Overview on Higgs-dilaton cosmology}
\label{subsec:Overview_on_Higgs-dilaton_cosmology}

We will now proceed to describe the main features of the Higgs-dilaton model.

\begin{itemize}
	\item \textit{Inflationary Era:} If the initial conditions of the scalar fields are far away from the potential valleys, they will roll slowly towards one of them. $\Lambda_0$ can be safely neglected as it is yet very small. This slow-roll of the fields is responsible for inflation, which, being driven by the Higgs field, is much like in the case of the Higgs-Inflation model of Ref.~\cite{Bezrukov:2007ep}. The era of inflation is eventually terminated by the preheating and reheating phases described in Ref.~\cite{GarciaBellido:2008ab}.
	\item \textit{Preheating Phase:} After inflation, the scalar field dynamics is dominated by the field $h$. The gauge bosons created at the minimum of the potential acquire a large mass and starts to decay into all of the SM particles. The fraction of energy going into Standard Model particles is yet small, so non-perturbative decay is slow (see in Ref.~\cite{GarciaBellido:2008ab}).
	\item \textit{Reheating phase:} The reheating of the Universe in the Higgs-inflation paradigm was been considered first in Ref.~\cite{Bezrukov:2008ut,GarciaBellido:2008ab}. Specifically, after some time, the amplitude of the oscillations becomes small enough  that the gauge boson masses become too small to induce a quick decay, and their occupation numbers start to grow rapidly via parametric resonance. Then the gauge bosons back-react on the Higgs field and preheating ends. From there on, the Higgs field as well as the gauge fields decay perturbatively until their energy is transferred to SM leptons and quarks.
	\item \textit{dark energy Era:} After preheating and reheating the scalar fields satisfy Eq.~\eqref{eq:Ground_States_EF_beta_Si_Cero} and remain virtually static for most of the present Universe history. The energy contained in the $h(t)$ and $\chi(t)$ fields stays almost unchanged, fixed at a given value by $\Lambda_0$. It does, however, eventually dominate the energy budget of the Universe as it scales with $\OL \propto a^{0}$ while radiation and matter energy densities scale with $\Or \propto a^{-4}$ and $\Om \propto a^{-3}$ respectively.\footnote{The density parameter for any cosmological perfect fluid ``$i$'' is defined as $\Omega_i = \rho_i / \rho_c$ where $\rho_c = 8 \pi G / 3 H^2$ is the critical density.}
	\end{itemize}

The next two sections will now describe the HDM phenomenology of the inflation and dark energy era in detail.

\section{HDM Implications for the early Universe}
\label{sec:HDM_Implications_for_the_early_Universe}

This section deals with the HDM predictions for the inflationary era and it is organized as follows: We first find the field trajectories after some suitable frame changes and field redefinitions. Then we proceed to calculate the scalar $\mathcal{P}_s(k)$ and tensor $\mathcal{P}_t(k)$ power spectra of perturbations and comment as well as on a number of interesting predictions, such as the absence of isocurvature perturbations.

It is thought that, during inflation, all the energy of the Universe was contained in the inflaton and gravitational fields, so one can readily neglect the Standard Model fields in the Lagrangian (see Ref.~\cite{GarciaBellido:1995qq}), $\tilde{{\cal L}}_{\rm{SM}[\lambda \rightarrow 0]} \rightarrow 0$. In the Einstein frame, as seen in Eq.~\eqref{eq:Lagrangian_EF}, then one is left with the scalar-tensor part of Eq.~\eqref{eq:Lagrangian_UG}
	\begin{equation}
	\label{eq:Lagrangian_EF_bis}
	\frac{\mathcal{L}}{\sqrt{-\tilde{g}}} = \frac{M_P^2}{2} \tilde{R} - \frac{1}{2} \tilde{K} - \tilde{U}(\boldsymbol{\phi})
	\;,
	\end{equation}
where we express the Higgs and dilaton fields in a vector manner for brevity, $\boldsymbol{\phi} = (\phi_1,\phi_2) = (h, \chi)$.

The kinetic and potential terms are described in Eqs.~\eqref{eq:Kinnetic_EF} and~\eqref{eq:Potential_EF_beta_Si_Cero} respectively, and the latter can be divided as the the sum of a scale-invariant term function of $V(\boldsymbol{\phi})$, and a scale-breaking term function of $\Lambda_0$:
	\begin{equation}
	\label{eq:Potential_EF}
	\tilde U(\boldsymbol{\phi}) = \tilde V(\boldsymbol{\phi}) + \tilde V_{\Lambda_0}(\boldsymbol{\phi})
	\;.
	\end{equation}

This arrangement allows to write the equations of motion for the gravitational (Einstein equations) and scalar (Klein-Gordon equation) fields as Eq.~\eqref{eq:Lagrangian_EF_bis},
	\begin{align}
	\label{eq:Einstein_EF}
	\begin{split}
	\tilde G_{\mu \nu} = 	& \tilde{U} \tilde{g}_{\mu \nu} + \gamma_{a b} \left(\partial_\mu \phi^a \partial_\nu \phi^b \right)
									\\
									& - \gamma_{a b} \left( \frac{1}{2} \tilde g_{\mu \nu} \tilde g^{\rho \sigma} \partial_\rho \phi^a \partial_\sigma \phi^b \right)
	\end{split}					\\
	\label{eq:Klein-Gordon_EF}
	\tilde U^{;c} =				& \tilde{\Box} \phi^c + \tilde g^{\mu \nu} \Gamma^c_{ab} \partial_\mu \phi^a \partial_\nu \phi^b
									\;,
	\end{align}
where the Einstein tensor $\tilde G_{\mu \nu}$ is computed from the newly conformal metric $\tilde{g}_{\mu \nu}$, and $\Gamma^c_{a b}$ are the Christoffel symbols corresponding to the space metric $\gamma_{a b}$.

Since inflation only takes place in the scale invariant region of the potential, one can neglect the scale invariance breaking term $\Lambda_0$. In the Einstein frame, as the metric $\tilde g_{\mu\nu}$ is scale invariant by definition, scale transformations do not act on it and thus ${\delta \tilde{g}_{\mu \nu} = 0}$, but only affect the scalar fields. This allows to redefine the fields as $\boldsymbol{\phi}' = ({\phi'}^1,{\phi'}^2) = (\rho,\theta)$, which in terms of the original fields $\boldsymbol{\phi} = ({\phi}^1, {\phi}^2) = (h,\chi)$ (see Ref.~\cite{GarciaBellido:2011de}), are
	\begin{align}
	\label{eq:Rho_Definition}
	\rho = 	& \frac{M_P}{2} \ln \left( \frac{(1 + 6 \xc ) \chi^2 + (1 + 6 \xh ) h^2}{M_P^2} \right)
				\\
	\label{eq:Theta_Definition}
	\theta = 	& \arctan\left(\sqrt{\frac{1 + 6\xh}{1 + 6\xc}}\frac{h}{\chi}\right)
				\;,
	\end{align}
so that the scale transformation only acts on a single field, $\rho$.

Note that the argument of the logarithm in Eq.~\eqref{eq:Rho_Definition} is reminiscent of the two-dimensional ellipse equation in the $(h,\chi)$ plane and can be interpreted as the radius $\rho$ of the field vector $\boldsymbol{\phi}$, while from the argument of the arctangent in Eq.~\eqref{eq:Theta_Definition} is clear that $\theta$ can be interpreted as the the argument of the field vector $\phi$. One may tentatively refer to this new variables as ``polar fields'' and we will do so throughout the rest of the paper.

In terms of the redefined fields, the Einstein frame kinetic term $\tilde K$ and the potential $\tilde U = \tilde V + \tilde V_{\Lambda_0}$ [see Eq.~\eqref{eq:Lagrangian_EF_bis} are given by
	\begin{align}
	\label{eq:Kinetic_Polar}
	\begin{split}
	\tilde K(\theta,\partial \rho,\partial \theta) = 	& \left( \frac{1 + 6 \xh}{\xh}\right)\frac{1}{\sin^2 \theta + \varsigma \cos^2 \theta} \left( \partial \rho \right)^2
																	\\
																	& + \frac{M^2\, \varsigma}{\xc} \frac{\tan^2 \theta + \mu}{\cos^2 \theta \left(\tan^2 \theta + \varsigma \right)^2} \left(\partial \theta \right)^2
	\end{split} 													\\
	\label{eq:Potential_Si_SI_Polar}
	\tilde V(\theta) = 											& \frac{\lambda M_P^4}{4\xh^2}\left(\frac{\sin^2 \theta-\frac{\alpha}{\lambda}\frac{1 + 6 \xh}{1 + 6 \xc}\cos^2 \theta}{\sin^2 \theta + \varsigma \cos^2 \theta} \right)^2\
																	\\
	\label{eq:Potential_No_SI_Polar}
	\tilde V_{\Lambda_0}(\rho,\theta) = 				& \left( \frac{1 + 6 \xh}{\xh}\right)^2 \frac{\Lambda_0 e^{-4 \rho / M_P}}{\left (\sin^2 \theta + \varsigma \cos^2 \theta \right)^2}
																	\,,
	\end{align}
where $\mu \equiv \xc / \xh$ and $\varsigma \equiv (\xc + 6 \xh \xc ) ( \xh + 6 \xc \xh )$.

\subsection{Evolution equations for the background}
\label{subsec:Evolution_equations_for_the_background}

We now proceed to study the field trajectories during inflation in order to link the couplings of the theory with inflationary quantities. In a Friedmann-Lema\^{i}tre-Robertson-Walker (FLRW) background,
	\begin{equation}
	ds^2 = \tilde g_{\mu\nu}dx^\mu dx^\nu = -dt^2 + a^2(t)d\vec{x}^2
	\;,
	\end{equation}
and following Ref.~\cite{GarciaBellido:2011de}, the Friedmann and Klein-Gordon equations are\footnote{A dot $\dot{}$ denotes a derivative w.r.t. time.}\footnote{Capital $D$ stands for the usual covariant derivative.}\footnote{A dagger $\dagger$ indicates the dual of a vector.}
	\begin{align}
	\label{eq:Friedmann_1_Time}
	H^2 									& = \frac{1}{3 M_P^2} \left( \frac{1}{2} \big| \dot{\boldsymbol{\phi}} \big|^2 + \tilde{U} \right)
											\\
	\label{eq:Friedmann_2_Time}
	2 \dot{H} + 3 H^2 				& = - \frac{1}{M_P^2} \left( \frac{1}{2} \big| \dot{\boldsymbol{\phi}} \big|^2 - \tilde{U} \right)
											\\
	\label{eq:Klein-Gordon_Time}
	- \nabla^\dagger \tilde{U} 	& = \frac{D \dot{\boldsymbol{\phi}}}{dt} + 3 H \dot{\boldsymbol{\phi}}
											\;,
	\end{align}
where $H = \dot{a}/a$ is the Hubble rate.

If, instead of the time coordinate $t$, we use the e-fold parameter $N = \ln a(t)$, the field equations can then be rewritten as\footnote{A prime $'$ denotes a derivative w.r.t the number of e-folds.}
	\begin{align}
	\label{eq:Friedmann_1_e-Folds}
	H^2 =												& \frac{\tilde U}{3 M_P^2 - \big| \boldsymbol{\phi}' \big|^2 / 2}
															\\
	\label{eq:Friedmann_2_e-Folds}
	\frac{H'}{H} = 										& -\frac{1}{2}\frac{\big|\boldsymbol{\phi}'\big|^2}{M_P^2}
															\\
	\label{eq:Klein-Gordon_e-Folds}
	- M_P^2 \nabla^\dagger \ln \tilde U = 	& \boldsymbol{\phi}' + \frac{ D \boldsymbol{\phi}' / dN}{3 - \big| \boldsymbol{\phi}' \big|^2 / 2 M_P^2}
															\;.
	\end{align}

We will solve this equation in the slow-roll (SR) formalism.

\subsubsection{Slow-roll parameters}
\label{subsubsec:Slow-roll_Parameters}

In the HDM, inflation occurs due to a phase of slow roll of the scalar fields over an almost-flat potential long before the fields reach one of the potential valleys,. It is then convenient (see Ref.~\cite{Lyth:1993eu,Peterson:2010np}) to define the slow-roll parameter $\epsilon^{SR}$ and the slow-roll vector $\boldsymbol{\eta}^{SR}$ as
	\begin{align}
	\label{eq:Slow-Roll_epsilon}
	\epsilon^{SR} 				& \equiv - \frac{H'}{H} = \frac{1}{2} \frac{\big| \boldsymbol{\phi}' \big|^2}{M_P^2}
										\\
	\label{eq:Slow-Roll_eta}
	\boldsymbol{\eta}^{SR} 	& \equiv \frac{1}{ | \boldsymbol{\phi}' \big|}\frac{D \boldsymbol{\phi}'}{dN}
										\;,
	\end{align}
where the vector notation is introduced due to the presence of more than one field in the model.

For inflation to take place, the slow roll conditions $\epsilon^{SR} \ll 1$ and $| \boldsymbol{\eta}^{SR}| \ll 1$ must be satisfied\footnote{Parameters $\epsilon^{SR} = \epsilon^{SR}_1$ and $|\boldsymbol{\eta}^{SR}| = |\epsilon^{SR}_2| / 2$, where $\epsilon^{SR}_1\equiv H' / H$ and $\epsilon^{SR}_2\equiv d \ln \epsilon^{SR} /dN$ are the multifield generalizations of the standard first two horizon-flow parameters defined in Ref.~\cite{Schwarz:2001vv}.}. Inflation ends when the slow-roll parameters cease to be smaller than unity or a phase transition occurs.

Imposing the slow-roll conditions, and in terms of of the number of e-folds, the Friedmann Equations~\eqref{eq:Friedmann_1_e-Folds},~\eqref{eq:Friedmann_2_e-Folds} and the Klein-Gordon equation~\eqref{eq:Klein-Gordon_e-Folds} can now be rewritten as:
	\begin{align}
	\label{eq:Friedmann_Slow-Roll}
	H^2 						&= \frac{\tilde U}{3M_P^2}
								\\
	\label{eq:Klein-Gordon_Slow-Roll}
	\boldsymbol{\phi}' 	&= -M_P^2 \nabla^\dagger\ln\tilde U
								\;,
	\end{align}
giving the evolution of the field vector $\boldsymbol{\phi}$.

\begin{figure}[!t]
\centering
\vspace{0cm}\rotatebox{0}{\vspace{0cm}\hspace{0cm}\resizebox{0.45\textwidth}{!}{\includegraphics{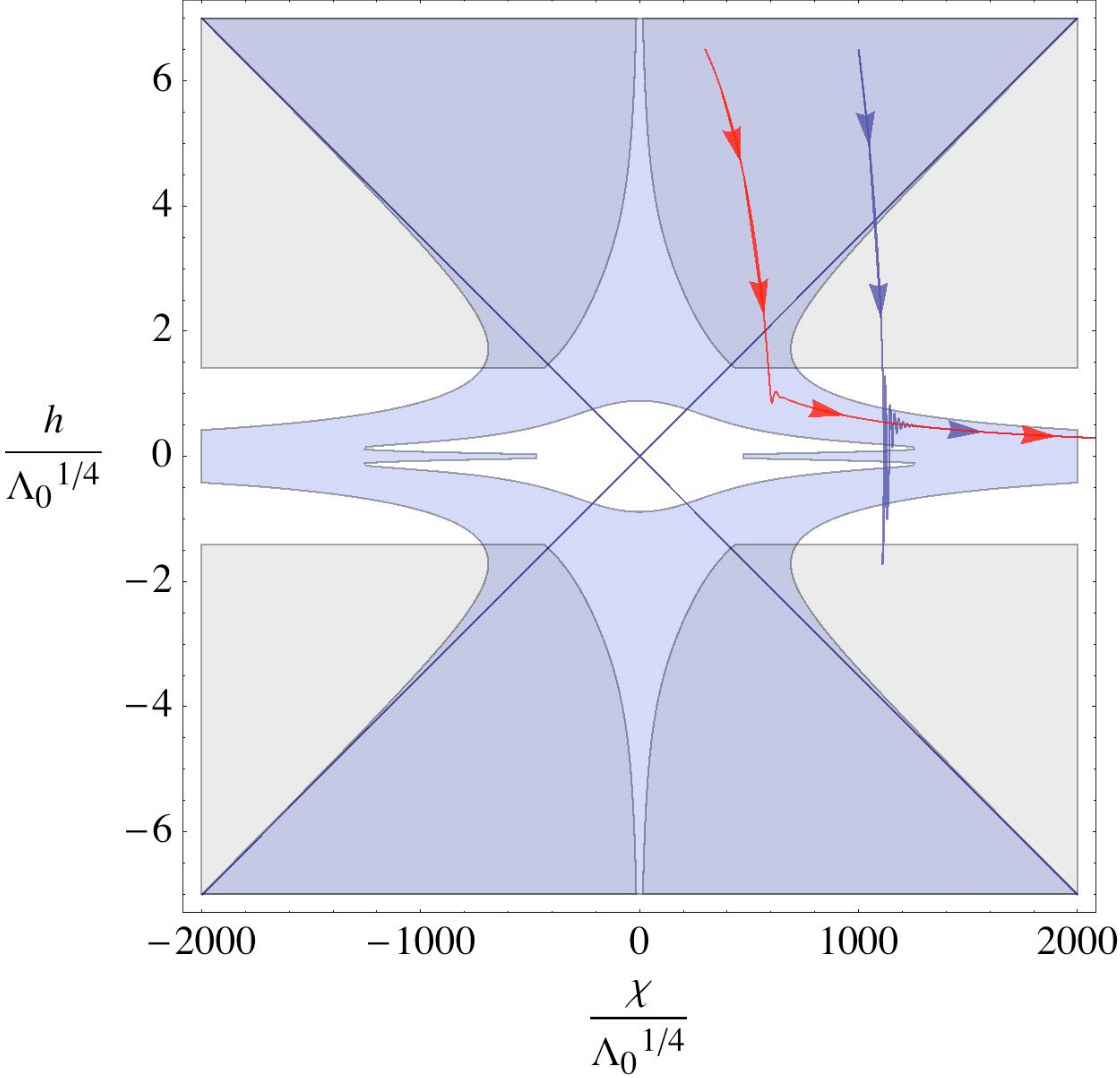}}}
\caption{The blue region is the slow-roll region for $\xc\ll 1$, $\xh\gg 1$ and the shaded region is the scale-invariant region. The red line represents a trajectory, that never leaving the slow-roll region, thus never producing a preheating/reheting phase, while the blue line is a trajectory that leaves the slow-roll region and oscillates strongly before rolling down the potential valley giving rise to a preheating/reheting phase. The Figure is taken from Ref.~\cite{GarciaBellido:2011de}.}
\label{fig:Slow-Roll_Map}
\end{figure}

\subsubsection{Background trajectories}
\label{subsubsec:Background_trajectories}

Lets discuss the regions in the $(h,\chi)$ plane for which the approximate slow-roll conditions hold (see Fig.~\ref{fig:Slow-Roll_Map}). Note that the smallness of $\alpha$ results in that the potential valley do barely separate from each other at late times and are indeed merged for the whole period of inflation. Hence, for the rest of this section $\alpha = 0$ will be assumed.

The initial conditions for the trajectories must be chosen in the slow-roll region; the shape of the potential Eq.~\eqref{eq:Potential_EF} attracts all trajectories to one of the potential valleys, and the roll starts. Note that:
	\begin{itemize}
	\item Some trajectories never leave the slow-roll sector and oscillate only a small number of times before falling down the potential valley. Such trajectories do not go through a reheating phase (see the red line in Fig.~\ref{fig:Slow-Roll_Map}).
	\item Other trajectories do leave the slow-roll sector by reaching a stepper region within the potential valley and oscillate strongly around its minimum allowing for the reheating phase to take place before settling in the potential valley (see the blue line in Fig.~\ref{fig:Slow-Roll_Map}).
	\end{itemize}

As reheating is a necessary component of any viable theory of inflation, it is the second class of trajectories to which we adhere for the rest of the analysis, constraining the initial conditions of the fields. For these trajectories, and in terms of the polar redefinition of the fields $(\rho,\theta)$ of Eqs.~\eqref{eq:Rho_Definition} and~\eqref{eq:Theta_Definition}, the slow-roll equations for the scalar fields Eq.~\eqref{eq:Klein-Gordon_Slow-Roll} become
	\begin{align}
	\label{eq:Rho_Slow-Roll+SI}
	\rho' =		& 0
					\\
	\label{eq:Theta_Slow-Roll+SI}
	\theta' = 	& -\frac{4 \xc}{1 + 6 \xc} \cot \theta \left( 1 + \frac{6 \xc \xh}{\kappa(\theta)} \right)
					\,.
	\end{align}
where $\kappa(\theta) = \xc \cos^2 \theta + \xh\ \sin^2 \theta$.

As a consequence of scale-invariance, the second scalar field equation Eq.~\eqref{eq:Theta_Slow-Roll+SI} does not depend on $\rho_0$ but merely on $\theta$. This ensures that there are no entropy perturbations in this model that may produce isocurvature fluctuations at reentry~\cite{GarciaBellido:1995qq}, which is quite positive for the HDM model since they are strongly constrained by \textit{Planck} satellite (2015 data release)~\cite{Planck:2013jfk}.

Integrating Eq.~\eqref{eq:Theta_Slow-Roll+SI} up to the end of inflation, that is, from $\theta$ to $\theta_\textnormal{end}$, the number of e-folds is
	\begin{equation}
	\label{eq:e-Folds_Theta}
	\begin{split}
	N(\theta,\theta_\textnormal{end}) = 	& \frac{1}{4 \xc} \ln \left( \frac{\cos \theta_\textnormal{end}}{\cos \theta} \right)
	 													\\
	 													& + \frac{3}{4} \ln \left(\frac{\kappa(\theta_{\textnormal{end}}) + 6 \xc \xh}{\kappa(\theta) + 6 \xc \xh} \right )
	 													\;,
	\end{split}
	\end{equation}
where $\theta_\textnormal{end}$ can be determined by finding its value when $\epsilon^{SR}(\theta_\textnormal{end}) = 1$,
	\begin{equation}
	\label{eq:e-Folds_Theta_End}
	\epsilon^{SR}(\theta_\textnormal{end}) = \frac{8 \xc^2 ( 1 + 6 \xh)}{1 + 6 \xc} \frac{\cot^2 \theta_\textnormal{end}}{\kappa(\theta_\textnormal{end})} = 1
	\;.
	\end{equation}

\subsection{Linear perturbations on the background}
\label{subsec:Linear_perturbations_on_the_background}

We will now proceed to calculate the scalar $\mathcal{P}_s(k)$ and tensor $\mathcal{P}_t(k)$ power spectra of perturbations from the evolution of the Higgs and dilaton fields, $h$ and $\chi$. This evolution is coded in the slow-roll parameters $\epsilon^{SR}$ and $\boldsymbol{\eta}^{SR}$, which ultimately depend on the theory couplings $\xh/\sqrt{\lambda}$ and $\xc$.

We make use of the theory of cosmological perturbations emerging from quantum fluctuations during inflation, developed, among others, by Ref.~\cite{Mukhanov:1990me}. Including scalar and tensor perturbations, and choosing the Newtonian transverse traceless gauge, the metric can be expanded in scalar (curvature) and tensor (gravitational waves) perturbations,
	\begin{align}
	\label{eq:Newtonian_Traceless_Gauge}
	\begin{split}
	ds^2 = 	& -\left(1 + 2 \Phi \right) dt^2
				\\
	 			& + a(t)^2 \left( \left( 1- 2 \Psi \right) \delta_{i j} + h_{i j} \right) dx^i dx^j
				\;,
	\end{split}
	\end{align}
where $\Phi$ and $\Psi$ are the Bardeen potentials of Ref.~\cite{Bardeen:1980kt}. Vector perturbations (vorticity) are not considered since they decay rapidly in any case.

\subsubsection{Primordial scalar power spectra}
\label{subsubsec:Primordial_scalar_power_spectra}

To compare the HDM with CMB observations, lets start with the power spectrum of the primordial scalar perturbations, (see Refs.~\cite{Bardeen:1980kt} and ~\cite{Kodama:1985bj}). The comoving curvature perturbation is
	\begin{equation}
	\label{eq:Curvature_Perturbation_Definition}
	 \zeta \equiv \Psi - \frac{H}{\dot H}\left(\dot \Psi + H \Phi \right)
	 \;.
	\end{equation}
It can be proven that $\zeta$ is conserved outside the horizon if inflation takes place in the scale-invariant region, as in this case just like in the single-field inflation scenario (see Refs.~\cite{GarciaBellido:1995fz} and ~\cite{GarciaBellido:1995qq}).

By making use of the slow-roll formalism,  it can be shown that the amplitude $A_s(k)$ of the scalar power spectrum $ \mathcal{P}_s(k) = A_s (k/k_0)^{\ns - 1}$ can be expressed as (see Ref.~\cite{Mukhanov:1990me})
	\begin{equation}
	\label{eq:s_Power_Spectrum}
	A_s(k) \simeq \frac{1}{2 M_P^2 \epsilon^{SR}} \left( \frac{H^{*}}{2 \pi} \right)^2 \simeq \frac{\tilde{V}^*}{24 \pi^2 \epsilon^{SR} M_P^4}
	\;,
	\end{equation}
where quantities with an asterisk are evaluated at the moment of horizon crossing, that is, when $aH = k_0$.

The scalar spectral index $\ns(k)$ is given by\footnote{For a formal definition of the speed-up rate $\eta^{SR}_\parallel$ and the turn rate $\eta^{SR}_\perp$, see Ref.~\cite{GarciaBellido:2011de}. They are the components of the slow-roll parameter $\boldsymbol{\eta}^{SR}$ in the parallel and perpendicular directions of the field trajectory, which define an orthonormal coordinate system of basis elements $e_\parallel$ and $e_\perp$.}
	\begin{equation}
	\label{eq:s-Index_Slow-Roll}
	 \ns(k) \equiv 1 + \frac{d \ln \mathcal{P}_s}{d \ln k} \simeq 1 - 2 (\epsilon^{SR} + \eta^{SR}_\|)
	 \;,
	\end{equation}
and the running of the spectral index $\as(k)$ can be expressed as
	\begin{equation}
	\label{eq:s-Running_Slow-Roll}
	\as \equiv \frac{d \ns}{d \ln k} \simeq -4\epsilon^{SR}\eta^{SR}_\|-2\eta^{SR}_\| \frac{d^2\ln \epsilon^{SR}}{dN^{2}}
	\;,
	\end{equation}
thus giving a non zero running.

\subsubsection{Primordial tensor power spectra}
\label{sebsebsec:Primordial_tensor_power_spectra}

Again by making use of the slow-roll approximation (see Ref.~\cite{Mukhanov:1990me}), the amplitude $A_t(k)$ of the power spectrum of the primordial tensor perturbations $\mathcal{P}_t(k) = A_t (k/k_0)^{n_t}$ can be written as
	\begin{equation}
	\label{eq:t_Power_Spectrum}
	A_t(k) \simeq \frac{8}{M_P^2} \left( \frac{H^{*}}{2 \pi} \right)^2 \simeq \frac{2 \tilde{V}^*}{3 \pi^2 M_P^4}
	\;,
	\end{equation}
which gives a tensor spectral index $n_t(k)$
	\begin{equation}
	\label{eq:t-Index_Slow-Roll}
	n_t(k) \equiv \frac{d \ln \mathcal{P}_t}{d \ln k} \simeq - 2 \epsilon^{SR}
	\;,
	\end{equation}
while we neglect the running of the tensor spectral index $\alpha_t = d n_t /d \ln k$.

Finally, it can be easily seen that the ratio of the tensor and the scalar spectra to first order in slow-roll is given by
	\begin{equation}
	\label{eq:ts_Ratio_SR}
	r \equiv \frac{A_t}{A_s} \simeq 16 \epsilon^{SR}
	\,
	\end{equation}
and we also have a consistency condition just like the one in the case of single-field inflation
	\begin{equation}
	\label{eq:Consistency_Relation}
	r = -8 n_t
	\,,
	\end{equation}
a relation which indeed holds for the vast majority of inflationary models in the slow-roll approximation at least to the first non-trivial order.

\subsection{CMB constraints on parameters and predictions}
\label{subsec:CMB_constraints_on_Parameters_and_predictions}

In this subsection we relate the primordial spectra observables from the CMB with the model couplings. Since the whole period of observable inflation takes place in the scale-invariant region of the potential, we can therefore use the background trajectories Eqs.~\eqref{eq:e-Folds_Theta} and~\eqref{eq:e-Folds_Theta_End} and directly compare to the primordial spectra calculated in Eqs.~\eqref{eq:s_Power_Spectrum} and~\eqref{eq:t_Power_Spectrum}, indirectly measured from the CMB, while assuming that during inflation, entropy (isocurvature) perturbations are not even excited, thanks to scale invariance of Eq.~\cite{GarciaBellido:2011de}.

Lets start by computing the spectral quantities $\mathcal{P}_s(k_0)$, $\ns(k_0)$, $\alpha(k_0)$ and $r(k_0)$ evaluated at the pivot scale $k_0$, in terms of the HDM couplings $\xc$, $\xh$ and $\lambda$. This will be done in the next four steps:
	\begin{itemize}
	\item \textit{Step I.} First, we start by parametrically solving Eq.~\eqref{eq:e-Folds_Theta_End}, giving the final state of the field $\theta_{end}$ at the end of the inflationary period, as a function of the HDM couplings.
	\item \textit{Step II.} Second, the field $\theta_{end}$ is inserted into Eq.~\eqref{eq:e-Folds_Theta}, parametrically solving as a function of the HDM couplings and obtaining $\theta^*$, the state of the field at the moment where the modes $k_0$ exits the horizon during inflation.
	\item \textit{Step III.} Third, the spectral quantities in Eqs.~\eqref{eq:s_Power_Spectrum},~\eqref{eq:s-Index_Slow-Roll},~\eqref{eq:s-Running_Slow-Roll} and~\eqref{eq:Consistency_Relation} are evaluated at $\theta^*$ to find the spectral quantities $P_s(k_0)$, $\ns(k_0)$, $\as(k_0)$ and $r(k_0)$, as functions of the HDM couplings and $N^*$, the number of e-folds between the moment where the modes $k_0$ exits the horizon and the end of inflation.
	\item \textit{Step IV.} Last, $N^*$ is expressed as a function of the HDM couplings, in order to check whether the model is able to provide with a number of e-folds large enough that the horizon, homogeneity and relic problems of $w_0w_a$CDM are solved.
	\end{itemize}

Before we start, however, it can be shown that the number of e-folds roughly corresponds to (see Ref.~\cite{Mather:1998gm})
\begin{equation}
\label{eq:e-Folds_Reheating}
\begin{split}
	N^* \simeq 	& 59 -\ln \frac{k_0 \mathrm{Mpc}}{0.002 a_0} - \ln \frac{10^{16} \,\mathrm{GeV}}{\tilde V(\theta^*)^{1/4}}
						\\
	 					& + \ln \left( \frac{\tilde V(\theta^*)}{\tilde V(\theta_{end})} \right)^{1/4} -\frac{1}{3} \ln \left( \frac{\tilde V(\theta_{end})}{\varrho_{\mathrm{rh}}} \right)^{1/4}
	 					\,,
\end{split}
\end{equation}
where of $\varrho_{\mathrm{rh}}$ is the energy density at reheating. It has been found in Ref.~\cite{GarciaBellido:2008ab} that reheating in Higgs inflation is very efficient, and henceforth we take the approximation that a negligible number of e-folds occur during the reheating phase -- that is, we consider the case on instantaneous reheating of Ref.~\cite{GarciaBellido:2011de} --, at $\theta_{end}$. Then,
\begin{align}
\label{eq:Reheating_Density}
	\varrho_{\mathrm{rh}}= \tilde V(\theta_{end})
	\,,
\end{align}
and one thus has $\xc \lesssim 10^{-3}$ and $\xh/sqrt{\lambda} \sim \Order(10^4)$. With this bounds in mind, one can safely neglect second-order terms in $\xc$, $1/\xh$ and as will latter be seen in Eq.~\eqref{eq:e-Folds} $1/N^*$.

Moving on the explicit computation of the spectral parameters in terms of this couplings, lets follow the above four steps.

\paragraph{Step I.}
Solving Eq.~\eqref{eq:e-Folds_Theta_End} up to first-order in $\xc$ and $1/\xh$, one obtains that the state of the field $\theta$ at the end of the inflationary period is
\begin{equation}
\label{eq:Theta_End}
\theta_{end} = 2\times 3^{\frac{1}{4}}\sqrt{\xc} \;.
\end{equation}

\paragraph{Step II.}
Solving Eq.~\eqref{eq:e-Folds_Theta} up to first-order in $\xc$ and $1/\xh$, the state of the field $\theta$, as a function of the number of e-folds at the moment of horizon crossing, is
\begin{equation}
\label{eq:Theta_Star}
\theta^*\simeq \arccos\left(e^{-4\xc N^*}\right) \;.
\end{equation}

\paragraph{Step III.}
To evaluate the spectral observables $A_s(k_0)$, $\ns(k_0)$ and $\as(k_0)$ at $\theta^*$, we insert Eq.~\eqref{eq:Theta_Star} into the definitions of the spectral quantities given in Eqs.~\eqref{eq:s_Power_Spectrum},~\eqref{eq:s-Index_Slow-Roll} and~\eqref{eq:s-Running_Slow-Roll}. We obtain
\begin{align}
\label{eq:s_Amplitude_Couplings}
A_s(k_0) & \simeq \frac{\lambda}{1152\pi^2\xc^2\xh^2} \sinh^2 \left(4\xc N^*\right)
\\
\label{eq:s-Index_Couplings}
\ns \left(k_0\right) & \simeq 1-8\xc\coth \left(4\xc N^*\right)
\\
\label{eq:s-Running_Couplings}
\as(k_0)&\simeq -32\xc^2 \csch^{2}\left(4\xc N^*\right)
\;.
\end{align}

As for the tensor-to-scalar ratio $r(k_0)$, recalling the consistency condition Eqs.~\eqref{eq:Consistency_Relation} and~\eqref{eq:ts_Ratio_SR}, one obtains
\begin{align}
\label{eq:ts_Ratio_Couplings}
r(k_0) = &-8n_t(k_0) \\
\simeq & 192\xc^2 \csch^{2}\left(4\xc N^*\right)
\,,
\end{align}
an interesting result, since one can see that in this approximation, $\as(k_0)$, $r(k_0)$ and $n_t(k_0)$ are related as
\begin{equation}
\label{eq:s_Ligature}
\as(k_0)\simeq-\frac{1}{6}\,r(k_0) = \frac{4}{3}\,n_t(k_0) \;,
\end{equation}
which can be interpreted as an approximate consistency condition for the HDM that it is not replicated in other inflationary theories.

\paragraph{Step IV.}
In order to find $N^*$ in terms of the parameters of the theory, we insert Eq.~\eqref{eq:Theta_Star} into Eq.~\eqref{eq:e-Folds_Reheating}, giving the approximate result
\begin{align}
\label{eq:e-Folds}
N^*&\simeq 64.5-\frac{1}{2}\ln{\frac{\xh}{\sqrt{\lambda}}}\approx 59
\,,
\end{align}
which is compatible with solving the homogeneity, horizon and relic problems of $w_0w_a$CDM and allows us to neglect second order terms in $1/N^*$ in Eqs.~\eqref{eq:s_Amplitude_Couplings}, ~\eqref{eq:s-Index_Couplings}, ~\eqref{eq:s-Running_Couplings}, and~\eqref{eq:ts_Ratio_Couplings}.

\subsubsection{A note on reheating}
\label{subsubsec:A_note_on_Reheating}

These results also provide limits on the reheating temperature $T_{\mathrm{rh}}$, defined as the initial temperature of the homogeneous radiation dominated universe. $T_{\mathrm{rh}}$ is related to $\varrho_{\mathrm{rh}}$ through
\begin{equation}
\label{eq:Reheating_Temperature}
\varrho_{\mathrm{rh}} = \frac{\pi^2}{30} g_{\rm eff}(T_{\mathrm{rh}})T_{\mathrm{rh}}^4
\,,
\end{equation}
where $g_{\rm eff}(T_{\mathrm{rh}}) =427/4 = 106.75$ is the effective number of relativistic degrees of freedom of the Standard Model. This translates into the following upper and lower bounds on the reheating temperature of (see Ref.~\cite{GarciaBellido:2011de})
\begin{equation}
\label{eq:Reheating_Temperature_Bounds}
\unit[10^{12}]{GeV} \lesssim T_{\mathrm{rh}} \lesssim \unit[10^{15}]{GeV}
\;.
\end{equation}

Note that this extra relativistic degree of freedom could in principle add to the effective number of light degrees of freedom in the Universe. This would affect the Big Bang nucleosynthesis (BBN) ratios for the observed element frequencies, and lay waste to one of the more accurate predictions of $w_0w_a$CDM. However, it can be shown that this is not the case since just after reheating ends the dilaton energy density is of order $\Order(10^{-7})$, i.e. virtually negligible, so it no longer contributes to the effective number of light degrees of freedom as shown in Ref.~\cite{GarciaBellido:2012zu}.

Also, it should be noted that according to Eq.~(\ref{eq:Reheating_Temperature_Bounds}) even though there is a broad range of reheating temperatures, as was found in Ref.~\cite{GarciaBellido:2012zu} the dilation was never in thermal equilibrium in this interval, due to its negligible rate of interaction (due to its suppressed derivative coupling), so one should not include it in the sum of degrees of freedom.

\section{HDM Implications for the late Universe}
\label{subsec:HDM_Implications_for_the_late_Universe}

In this section we show that the dilaton is a good candidate for quintessence (QE), that is, a dynamical dark energy (DE) candidate, and provide with testable relations between CMB observables and the dark-energy equation of state parameter.

\subsection{The dilaton as a source of dark energy}
\label{subsec:The_dilaton_as_a_source_of_dark energy}
After the phase of reheating, the system enters the radiation dominated stage, at the beginning of which the total energy density is given by $\varrho_{\mathrm{rh}}$, see Eq.~\eqref{eq:Reheating_Temperature}. At this moment the scalar fields have nearly settled down in one of two potential valleys:
\begin{equation}
\label{eq:Ground_States_DE}
h(t)^2 \simeq \frac{\alpha}{\lambda} \chi(t)^2
\,.
\end{equation}

One could further redefine the polar variables introduced in Eqs.~\eqref{eq:Rho_Definition} and~\eqref{eq:Theta_Definition} by
	\begin{align}
	\label{eq:Rho_Rescaling}
	\tilde \rho = 						& \gamma^{-1}\rho
											\\
	\label{eq:Theta_Rescaling}
	\tilde |\boldsymbol{\phi}'| = 	& |\boldsymbol{\phi}_0| - \frac{M_p}{a} \tanh^{-1} \left( \sqrt{1 - \varsigma} \cos \theta \right)
											\,,
	\end{align}
where have defined
	\begin{align}
	\label{eq:a_Definition}
	a 				& = \sqrt{\frac{\xc (1 - \zeta)}{\zeta}}
					\\
	\label{eq:gamma_Definition}
	\gamma 	& = \sqrt{\frac{\xc}{1 + 6 \xc}}
					\;.
	\end{align}

This new redefinition allows to express the symmetry-breaking approximate ground states, that is, the two potential valleys, as a simple function of $\tilde \theta$ and the scalar curvature perturbation $\zeta$
	\begin{equation}
	\label{eq:Ground_State_Rescaling}
	\tanh^2\left(\frac{a\,\tilde\theta(t)}{M_P}\right)\simeq\frac{1- \zeta}{1 + \frac{\alpha}{\lambda}\frac{1 + 6\xh}{1 + 6\xc}} = 1 - \zeta
	\,.
	\end{equation}

We assumed that the fields roll exactly at the bottom of the valleys -- the asymptotic solution -- even if we would naturally expect small deviations from this behaviour. In this case, $\tilde \theta$ is time-independent, and inserting the constraint ~\eqref{eq:Ground_State_Rescaling} in the Einstein frame Lagrangian ~\eqref{eq:Lagrangian_EF_bis}, one obtains
	\begin{equation}
	\label{eq:Lagrangian_Quintessence}
	\frac{\mathcal{L}}{\sqrt{- \tilde g}} \simeq \frac{M_P^2}{2} \tilde R-\frac{1}{2}(\partial \tilde \rho)^2 - \tilde V_{\mathrm{QE}}(\tilde \rho)
	\;,
	\end{equation}
where $V_{\mathrm{QE}}(\tilde \rho)$ is thawing quintessence-like potential, see Ref.~\cite{Shaposhnikov:2008xb}
	\begin{equation}
	\label{eq:Potential_QE}
	 \tilde V_{\mathrm{QE}}(\tilde \rho) = \frac{\Lambda_0}{\gamma^{4}} e^{- 4 \gamma \tilde \rho / M_P} \,,
	\end{equation}
of the run-away kind that allows the dilaton field to play the role of a dynamic dark energy. For an interesting discussion of thawing and freezing dark energy models and a comparison between various parametrizations see Ref.~\cite{Pantazis:2016nky}.

Lets now discuss in more detail the influence of the field $\tilde\rho$ on standard homogeneous cosmology. In the FLRW metric, the equation of motion for the homogeneous dynamic field $\tilde\rho(t)$ is given by
	\begin{equation}
	\label{eq:Klein-Gordon_QE_Original}
	 \ddot {\tilde\rho} + 3 H \dot{\tilde\rho} + \frac{d V_{\mathrm{QE}}}{d\tilde\rho} = 0
	 \;.
	\end{equation}
	
The equation of state parameter $w_i$ of any perfect fluid ``$i$'' is $w_i \equiv p_i / \varrho_i$, and so for the scalar field $\tilde\rho$ it is
	\begin{equation}
	\label{eq:QE_Parameter_Definition}
	\wQE \equiv \frac{p_{\mathrm{QE}}}{\varrho_{\mathrm{QE}}} \equiv \frac{\dot{\tilde \rho}^2 / 2 - V_{\mathrm{QE}}}{\dot{\tilde \rho}^2 / 2 + V_{\mathrm{QE}}}
	\,,
	\end{equation}
and thus the equation of motion of dark energy can be more compactly written as
	\begin{equation}
	\label{eq:Klein-Gordon_QE}
	\dot \varrho_{\mathrm{QE}} = - 3 H \varrho_{\mathrm{QE}} \left( 1 + \wQE \right)
	\,.
	\end{equation}

Also, for a barotropic fluid of energy density $\varrho_b$, the Hubble parameter is given by the first Friedmann equation as
	\begin{equation}
	\label{eq:Friedmann_QE}
	 H^2 = \frac{1}{3 M_P^2}\left( \varrho_b + \varrho_{\mathrm{QE}} \right)
	 \,,
	\end{equation}
which in terms of the relative abundances $\Omega_i = \varrho_i / 3 M_P^2 H^2$ for the perfect fluid ``$i$'' can be written as the cosmic sum rule ${\Omega_\mathrm{m} + \OQE = 1}$, for flat space, neglecting the radiation and neutrino contributions.

It is useful to rewrite Eqs.~\eqref{eq:Klein-Gordon_QE_Original} and~\eqref{eq:Friedmann_QE} in terms of the dark-energy energy density, $\OQE$, and deviation from pure cosmological constant parameter, ${\dQE \equiv 1 + \wQE}$. This allows us to write the scalar field evolution equation\eqref{eq:Klein-Gordon_QE} and Friedmann equation~\eqref{eq:Friedmann_QE} as
	\begin{align}
	\label{eq:Klein-Gordon_QE_rewrited}
	\dQE' 		& = (2-\dQE) \left( -3\dQE + 4\gamma\sqrt{3\dQE\OQE} 	 \right)
							\\
	\label{eq:Friedmann_QE_rewrited}
	\Omega'_{\mathrm{QE}} 	& = 3 (\delta_i -\dQE) (1 - \OQE) \OQE
							\;,
							.
	\end{align}

Defining ${\delta_i \equiv 1 + w_i}$, where the subindex ``$i$'' stands for any barotropic fluid, be it radiation ``$r$'' or matter ``$m$'', it follows immediately that
\begin{itemize}
	\item Radiation has $p_\mathrm{r} = \rho_\mathrm{r}/3 \Rightarrow w_\mathrm{r} = 1/3$, so $\delta_\mathrm{r} = 4/3$.
	\item Dust has $p_\mathrm{m} = 0 \Rightarrow w_\mathrm{m} = 0$, so $\delta_\mathrm{m} = 1$.
	\item A pure cosmological constant holds exactly that $p_\Lambda = - \rho_\Lambda \Rightarrow w_\Lambda = -1$, so $\delta_\Lambda = 0$.
	\item A more general dark energy component only needs $p_\mathrm{DE} < -\rho_\mathrm{DE}/3 \Rightarrow \wDE < -1/3$, so $\dDE < 2/3$.
\end{itemize}

In Refs.~\cite{Copeland:1997et} and ~\cite{Ferreira:1997hj} is shown that, for $0 \leq \delta_b \leq 2$, the field trajectories approach one of two different attractor solutions. Which of these two depends on the value of $\gamma$:
\begin{enumerate}
	\item If $4\gamma > \sqrt{3 \delta_i}$, the field evolution is driven towards a stable fixed point $\OQE = 3 \delta_i / 16 \gamma^2$ with $\dQE = \delta_i$. In this case, the scalar field can only account at best for a small contribution to dark-energy as it gives rise to a baryon-like equation of state parameter, $\wQE (\gamma > \sqrt{3 \delta_i} / 4) = 1$.
	\item If $4\gamma < \sqrt{3 \delta_i}$, the field evolution is driven towards a different stable fixed point $\OQE = 1$ with $\dQE = 16 \gamma^2 / 3$. the scalar field can in this case describe the late-time acceleration of the Universe since it develops an equation of state parameter $\wQE (\gamma < \sqrt{3 \delta_i}/4) = 16\gamma^2/3-1$, less than $-1/3$ if $\gamma < \sqrt{3\delta_i}/4$ or alternatively $\xc < 1/2$.
\end{enumerate}

It was found in Sec.~\ref{sec:HDM_Implications_for_the_early_Universe} that the Higgs-dilaton model is able to describe inflation as long as the coupling of the dilaton field to gravity is much smaller that unity, i.e. $\xc \lesssim 10^{-2}$ so trajectories are assured to approach the second attractor, and accelerated expansion of the Universe is bound to occur.

\subsection{Dark-energy constraints on parameters and predictions}
\label{subsec:Dark-energy_constraints_on_Parameters_and_redictions}

We now proceed to give the explicit dependence of the equation of state parameters and the couplings of the theory, just as we did in the previous section for the scalar and tensor power spectra. It was shown in subsection~\ref{subsec:Non-minimal_SM_coupling_to_Unimodular_Gravity} that for the case where the potential in Eq.~\eqref{eq:Potential_EF} had $\beta = 0$, all of the dark-energy would consist of the $\tilde\rho$ scalar field energy density and be analogous to quintessence, while for the case with $\beta \neq 0$, dark energy would have an additional contribution from the cosmological constant term, $\Lambda_0$. In this paper no distinction is being made, however, between quintessence and dark energy, as we are operating under the first assumption, and for the rest of the analysis we switch from ``QE'' to ``DE'' as the corresponding label for quintessence/dark energy observables.

The Inflationary era is followed by the Radiation and Matter dominated eras. During this epoch, the second term on the right-hand side of Eq.~\eqref{eq:Klein-Gordon_QE_rewrited} is small compared to the first one $\dDE$ is set at an equation of state parameter virtually indistinguishable from a pure cosmological constant, $\wDE \simeq -1$, yet of low energy density. However, since the energy density of quintessence barely decreases over time, $\ODE$ eventually becoming relevant. It is then when the scalar fields ${\tilde\rho}$ starts rolling faster down the potential valleys and $\dDE$ starts growing towards its attractor value, driving the accelerated expansion of space.

Note that while the Universe is not yet purely quintessence-dominated, $\dDE \ll 1$, Eqs.~\eqref{eq:Klein-Gordon_QE_rewrited} and~\eqref{eq:Friedmann_QE_rewrited} yield (for a detailed calculation, see Ref.~\cite{Scherrer:2007pu})
	\begin{equation}
	\label{eq:Big_Parameter_QE}
	\frac{3 \dDE}{16 \gamma^2} \simeq F^2
	\;,
	\end{equation}
where $F$ is the function defined below,
	\begin{equation}
	\label{eq:Big_Parameter_F}
	F(\ODE) \equiv \frac{\ODE + 2 \sqrt {\ODE} -1}{2 \ODE} \ln \frac{1 + \sqrt{\ODE}}{1-\sqrt{\ODE}}
	\;,
	\end{equation}
increasing from $F(0) = 0$ in the Radiation and Matter eras to $F(1) = 1$ when quintessence becomes fully dominant. Note that, in conjunction with Eq.~\eqref{eq:gamma_Definition}, giving the dependence of parameter $\gamma$ on the coupling of the dilaton $\xc$ to GR, we are at last able to link the inflationary quantities with dark energy observables, and it can be shown in particular that, for $\xc \lesssim 10^{-3}$, then $\dDE \lesssim 10^{-2}$ as well.

\subsection{Constrain relations between early and late Universe observables}
\label{subsec:Constrain_relations_between_early_and_late_Universe_observables}

We finally arrive to the point where three expressions, one linking first order parameters $\woDE$ and $\ns(k_0)$, another linking second order parameters $\waDE$ and $\as(k_0)$, and one last on $r(k_0)$ and $\woDE$ are deduced. These can be understood as consistency checks on the model, since they put very stringent bounds on the predicted values of $r(k_0)$ and $\woDE$ respectively.

\subsubsection{First-order consistency relations}
\label{subsubsec:First-order_consistency_relations}

We can test the constraint relation between the scalar spectral index $\ns$ and the equation of state parameter $\woDE$ as they both depend on $\xc$, for a given number of e-folds $\Ninf$. Lets then find the explicit relation between these magnitudes. Combining Eqs.~\eqref{eq:Big_Parameter_QE}, ~\eqref{eq:gamma_Definition} and~\eqref{eq:s-Index_Couplings} one may write the scalar tilt $\ns$ as a function the quintessence equation of state parameter $\dDE$ and the number of e-folds $\Ninf$.
	\begin{equation}
	\label{eq:1st_Order_Relation_Delta}
	\ns = 1- \frac{2}{\Ninf} G \coth G
	\;,
	\end{equation}
where the function $G(\ODE,\dDE)$ is defined as
	\begin{align}
	\label{eq:Big_Parameter_G}
	G(\OQE,\dDE, \Ninf) = \frac{6 \dDE \Ninf}{8 F(\OQE)-9 \dDE}
	\;,
	\end{align}
so a relationship between $\ns$ and $\woQE$ can be established, and it is enough to replace $\dDE = 1 + \wDE$ in Eq.~\eqref{eq:1st_Order_Relation_Delta}, which for the case where $\dDE \rightarrow 0$ reduces to:
	\begin{equation}
	\label{eq:1st_Order_Relation_Limit}
	\lim_{\dDE \rightarrow 0} \ns = \frac{\Ninf - 2}{\Ninf}
	\;.
	\end{equation}

Finally, we note that this result can be equivalently written in the differential expression
	\begin{equation}
	\label{eq:1st_Order_Relation_Derivative}
	\frac{d\ln\varrho_{\mathrm{QE}}}{d\ln a} \Bigg|_{a^*} \simeq \frac{d\ln P_s(k)}{d\ln k} \Bigg|_{k_0}
	\,,
	\end{equation}
at horizon reentry, that is, when $k_0 = a^*H$.

Note that the last equation, being a relation between the very small and the very big scales, implies a linear relation between the deviations from scale invariance in $\ns$ and the deviations from the pure cosmological constant dark energy equation of state parameter $\dDE$, which can be understood as a consequence of scale invariance.

\subsubsection{Second-order consistency relations}
\label{subsubsec:Second-order_consistency_relations}

We can also test a constraint relation relation involving second order quantities: the running of the scalar spectral index $\as$, the equation of state parameter $\woDE$, and its running with the scale factor, $waDE$, again as they both depend on $\xc$, for a given number of e-folds $\Ninf$.\footnote{We have chosen the following parametrization of the equation of state parameter for dark energy in terms of the scale factor: $\wDE (a) = \woDE + \waDE\ln(a/a_0)$. Note the evolution is logarithmic, not linear, which is taken into account when implementing the two consistency relations in \cosmomc.}
	\begin{equation}
	\label{eq:2nd_Order_Relation_Delta}
	\begin{split}
	\as = -\frac{8 \waDE F}{3 \dDE N^2_{\mathrm{inf}}} G^2 \left( \text{csch}^2 G - G \coth G \right)
	\;,
	\end{split}
	\end{equation}
which for the case where $\dDE \rightarrow 0$ reduces to
	\begin{equation}
	\label{eq:2nd_Order_Relation_Limit}
	\lim_{\dDE \rightarrow 0} \as = 0
	\;.
	\end{equation}
	
Along with Eq.~\eqref{eq:1st_Order_Relation_Delta}, this too can be considered a consistency check for the theory, albeit a second order one, whose testability however may still have to wait for a longer time than for the first-order case since they are as of yet very poorly constrained. Also, in a similar manner as with the relation for first order quantities we have the equivalent relation for the second order quantities
	\begin{equation}
	\label{eq:2nd_Order_Relation_Derivative}
	\frac{d^2 \ln \varrho_{\mathrm{DE}}}{(d \ln a)^2} \Bigg|_{a^*}\simeq \frac{d^2 \ln \mathcal{P}_s(k))}{(d \ln k)^2} \Bigg|_{k_0}
	\,,
	\end{equation}
at horizon reentry, that is, when $k_0 = a^*H$. This is again a non trivial result that, just as in the case of the first-order relation, may be understood as a consequence of scale invariance.

\subsubsection{Constraints on the tensor-to-scalar ratio}
\label{subsubsec:Constraints_on_the_tensor-to-scalar_ratio}

There is an additional constraint for the tensor-to-scalar ratio arising from Eq.~\eqref{eq:ts_Ratio_Couplings}, which, in terms of the energy density and equation of state parameters of dark energy reduces to:
	\begin{equation}
	\label{eq:r_Relation}
	r = \frac{12}{N^2_{\mathrm{inf}}} G^2 \text{csch}^2 G
	\;,
	\end{equation}
which for the case where $\dDE \rightarrow 0$ reduces to
	\begin{equation}
	\label{eq:r_Relation_Limit}
	\lim_{\dDE \rightarrow 0} r = \frac{12}{\Ninf^2}
	\;.
	\end{equation}

Note that the value of the tensor-to-scalar ratio $r$ will very rapidly go to zero for a sufficiently large number of e-folds, and indeed, considering the previously found value of $\Ninf \gtrsim 59$ and \textit{Planck} satellite (2015 data release) ~best fit values for parameters $\ODE$ and $\dDE$, it can be easily checked that the HDM prediction for the tensor-to-scalar ratio es indeed extremely constraining, at $r \lesssim 10^{-2}$, a very interesting feature of the theory that may be tested in the not-so-far future.


\section{Numerical results}
\label{sec:Numerical_results}

This section is organized as follows. First we review the codes used for the cosmological parameter space sampling, \cosmomc\mbox{} and \camb, and marginalization of the samples, \getdist\mbox{} and comment the modifications within these codes. Second, describe the cosmological probes whose likelihoods we consider in our runs: the \textit{BICEP}, \textit{Keck}, \textit{Planck}, Baryon Acoustic Oscillations (\textit{BAO}) and Joint Light-curve Analysis(\textit{JLA}). We include in our runs the lensing (\textit{len.}), low multipoles (\textit{low$\ell$}) and Matter Power spectrum (\textit{MPK}) likelihoods, and hereafter refer to the collection of likelihoods used as \textit{BKP+MPK+len.+low$\ell$+ext.}, explain our choice of prior distributions and other sampling details. Third, we present the histograms and contours of the correlated regions of a number of cosmological parameters for both $w_0w_a$CDM and HDM, and test the theoretical predictions of this parameters by sample selection in the vanilla $w_0w_a$CDM with those of the fully modified HDM \cosmomc\mbox{} computations. Fourth, we revert the constraints to find the Higgs-dilaton model predictions for the couplings. Finally, we compare the evolving dark energy cosmology of $w_0w_a$CDM to the HDM by Bayesian analysis, and comment on the likelihood and compatibility of each of these.

\subsection{\cosmomc\mbox{} modifications}
\label{subsec:CosmoMC_modifications}

The Markov chain Monte Carlo (MCMC) methods are belong to class of iterative algorithms that sample parameter space from a prior distribution on said parameters, based on constructing a Markov chain, that samples the desired posterior distribution. The quality of the sample improves with the number of steps. Eventually, after certain convergence criteria are met, the code can be terminated and one may marginalize the cosmological parameters from the chains, with the ``burn-in'' removed. At this point, if the chain has reached convergence well enough that the posterior distribution is fully sampled, then continuing to run the chain should have a negligible effect on the output statistics.

\cosmomc\mbox{} is a Fortran MCMC code for exploring cosmological parameter space, see Ref.~\cite{CosmoMC-Readme:2016}. The code does compute accurate theoretical matter and CMB power spectra with the help of the \camb\mbox{} program, see Ref.~\cite{CAMB-Readme:2016}, itself a Boltzmann code for CMB anisotropies. It then produces a set of chain files that includes the chain values of the cosmological parameters requested, and it comes together with a Python tool, \getdist\mbox{} (see Ref.~\cite{CosmoMC-Python:2016}), that analyzes the chain files, marginalizes the parameters and outputs the results in 1D, 2D and 3D posterior distributions, as well as providing marginalization, likelihood and convergence statistics.

\cosmomc\mbox{} comes with variety of sampling methods, from which we have selected the Metropolis-Hastings algorithm. In order to reach convergence faster, we have chosen \cosmomc\mbox{} to use an estimate of the covariance matrix for the \textit{Planck} satellite (2015 data release) data on some of the parameters and a fast/slow parametrization scheme (see Ref.~\cite{Lewis:2013hha}), reducing the computation time.

In all our runs, we do restrict ourselves to the vanilla version of the code for the $w_0w_a$CDM cosmology, but we had to modify the code in order to implement the logarithmic equation of state for dark energy $\wDE (a) = \woDE + \waDE\ln(a/a_0)$, the first order constraint $\ns \leftrightarrow (\woDE,\ODE,\Ninf)$, the second order constraint $\as \leftrightarrow (\ODE,\woDE,\waDE,\Ninf$ and the relation $r \leftrightarrow (\woDE,\ODE,\Ninf)$ of the HDM. We do so by modifying the \camb\mbox{} source files to calculate $\mathcal{P}_s(k)$ and $r(k)$ in terms of this newly defined spectral quantities, suitably defined within the source files and taking as input the values of $\ODE,\woDE,\waDE$ and $\Ninf$ from \cosmomc\mbox{} in each step of the chains.

For our runs \cosmomc\mbox{} works with as many as 29 parameters in the $w_0w_a$CDM run and 27 parameters in the HDM run, most of which are fed to \camb\mbox{} to produce the power spectra. For the the $w_0w_a$CDM run, the most relevant of these cosmological parameters are $\Omega_b h^2$, $\Omega_c h^2$, $100\theta_{\mathrm{MC}}$, $\tau_{\mathrm{re}}$, $A_s$, $\ns$, $\as$, $r_{0.05}$, $\woDE$, $\waDE$, while for the HDM run the number of e-folds $\Ninf$ may be added and both $\ns$, $\as$ and $r_{0.05}$ parameters removed from the MCMC, thus the difference of two in the number of degrees of freedom between the two models. Most of the remaining parameters ones are nuisance parameters and thus of no physical interest. Other parameters such as $\OK$, $\sum m_\nu$, $N_\nu$, $N_{\nu,m}$ and $N_{\nu,s}$ are fixed to their best available value or in the case of $n_t$ and $\alpha_t$ estimated to be zero, and are not fed into the MCMC.

It is worth noting that, in the HDM run, having added the spectral constraints to the code, neither $\ns$, $\as$ nor $r_{0.05}$ take part in the Metropolis-Hastings within \cosmomc, even if they do vary as derived functions of $\woDE$, $\waDE$, $\OL$ and $\Ninf$ inside \camb\mbox{} in each step. Theoretically, this would imply that no prior is needed in $\ns$, $\as$ and $r_{0.05}$ since they are ``dummy'' parameters in the initialization file. However, in practice, an input initial value for these parameters is needed for the code to start running properly. This choice of ``dummy'' priors are included in Tables~\ref{tab:Priors_CosmoMC} for completeness. Still, any other choice of said dummy priors does produce the exact same result, and in fact one may altogether set them of them to zero with no impact on the output.

\subsection{Details on the \cosmomc runs}
\label{subsec:Details_on_the_CosmoMC_runs}

	\begin{table*}[t!]
	\begin{tabular}{l | r@{}l r@{}l r@{}l r@{}l r@{}l }
	\hline \hline
	$w_0w_a$CDM 			& 		& $x_0$ 	& 		& $x_{min}$ 	& 		& $x_{max}$ 	& 		 & $\Delta_{max}$ 	 & 		& $\Delta_{prop}$ 	\\
	\hline
	$\Ob h^2$ 				& 0. 	& 022 		& 0. 	& 005 			& 0. 	& 100 			& 0. 	& 00010 				 & 0. 	& 00010 				\\
	$\Oc h^2$ 					& 0. 	& 120 		& 0. 	& 0010 			& 0. 	& 990 			& 0. 	 & 0010 					 & 0. 	& 0005 					\\
	$\theta_{MC}$ 			& 1. 	& 0411 		& 0. 	& 5000 			& 10. & 0000 			& 0. 	& 00040 				 & 0. 	& 00020 				\\
	$\tau_{RE}$ 				& 0. 	& 090 		& 0. 	& 010 			& 0. 	& 80 				& 0. 	 & 010 					 & 0. 	& 005 					\\
	$\log(10^{10} A_s)$ 	& 3. 	& 1 			& 2. 	& 0 				& 4. 	& 0 				 & 0. 	 & 0010 					 & 0. 	& 0010 					\\
	$\ns$ 						& 0. 	& 965 		& 0. 	& 915 			& 1. 	& 015 			& 0. 	 & 0020 					 & 0. 	& 0010 					\\
	$\as$ 						& -0. 	& 010 		& -0. & 035 			& 0. 	& 035 			& 0. 	 & 00028 				 & 0. 	& 00014 				\\
	$\woDE$ 					& -0. 	& 80 			& -1. & 80 				& 0. 	& 20 				 & 0. 	 & 0080 					 & 0. 	& 0040 					\\
	$\waDE$ 					& -0. 	& 40 			& -2. & 40 				& 1. 	& 60 				 & 0. 	 & 0160 					 & 0. 	& 008 					\\
	$r_{0.05}$ 					& 0. 	& 001 		& 0. 	& 000 			& 1. 	& 000 			& 0. 	 & 040 					 & 0. 	& 040 					\\
	$\Ninf$ 						& 		& n/a 		& 	& n/a 				& 		& n/a 			& 		 & n/a 					 & 		& n/a 					\\
	\hline \hline
	\end{tabular}
	\quad
	\begin{tabular}{l | r@{}l r@{}l r@{}l r@{}l r@{}l }
	\hline \hline
	HDM 						& 		& $x_0$ 	& 		& $x_{min}$ 	& 		& $x_{max}$ 	& 		 	 & $\Delta_{max}$ 	& 		& $\Delta_{prop}$ 	\\
	\hline
	$\Ob h^2$ 				& 0. 	& 022 		& 0. 	& 005 			& 0. 	& 100 			& 0. 	 	 & 00010 				 & 0. 	& 00010 				\\
	$\Oc h^2$ 					& 0. 	& 120 		& 0. 	& 0010 			& 0. 	& 990 			& 0. 		 & 0010 					 & 0. 	& 0005 					\\
	$\theta_{MC}$ 			& 1. 	& 0411 		& 0. 	& 5000 			& 10.	& 0000 			& 0. 	 	 & 00040 				 & 0. 	& 00020 				\\
	$\tau_{RE}$ 				& 0. 	& 090 		& 0. 	& 010 			& 0. 	& 80 				& 0. 		 & 010 					 & 0. 	& 005 					\\
	$\log(10^{10} A_s)$ 	& 3. 	& 1 			& 2. 	& 0 				& 4. 	& 0 				 & 0. 		 & 0010 					& 0. 	& 0010 					\\
	$\ns$ 						& 0. 	& 965 		& 0. 	& 915 			& 1. 	& 015 			& 0. 		 & 0020 					 & 0. 	& 0010 					\\
	$\as$ 						& -0. 	& 010 		& -0. 	& 035 			& 0.	& 035 			& 0. 	 	 & 00028 				 & 0. 	& 00014 				\\
	$\woDE$ 					& -1. 	& 00 			& -1. 	& 03 				& -0. 	& 97 				 & 0. 		 & 00030 				& 0. 	& 00015 				\\
	$\waDE$ 					& 0. 	& 00 			& -0. 	& 80 				& 0. 	& 80 				 & 0. 		 & 016 					& 0. 	& 008 					\\
	$r_{0.05}$ 					& 0. 	& 001 		& 0. 	& 001 			& 0. 	& 001 			& 0. 		 & 00010 				 & 0. 	& 00005 				\\
	$\Ninf$						& 62. & 0 			& 20. & 0 				& 240. & 0 			& 1. 	 	 & 1 						 & 2. 	& 2 						\\
	\hline \hline
	\end{tabular}
	\caption{Parameter priors for the main cosmological parameters for the $w_0w_a$CDM and HDM \cosmomc\mbox{} chains.}
	\label{tab:Priors_CosmoMC}
	\end{table*}
	
For the evaluation of both the $w_0w_a$CDM and HDM \cosmomc\mbox{} chains, we make use of the \textit{Bicep+}\textit{Keck+}\textit{Planck+}\textit{len.+}\textit{low$\ell$+}\textit{BAO+}\textit{JLA} likelihoods, provided in the July 2015 version of the code, the one used for for this paper's analysis.

Specifically, we have included the \textit{Planck} satellite (2015 data release) CMB likelihoods \cite{Aghanim:2015xee} and the gravitational lensing of the CMB from the trispectrum \cite{Ade:2015zua}, but also the joint analysis of data from the BICEP2/Keck Array \cite{Ade:2015tva}. In addition, we also include the ultraconservative cut of the galaxy weak lensing shear (WL) correlation function from the CFHTLenS survey \cite{Heymans:2013fya} and the matter power spectrum measurements from the clustering of the Sloan Digital Sky Survey DR7 Luminous Red Galaxies \cite{Reid:2009xm}.

We have also included distance data together with each of the perturbation-related data sets. Therefore, we have included the BAO measurements from CMASS and LOWZ of Ref.~\cite{Anderson:2013zyy}, the 6DF measurement from Ref.~\cite{Beutler:2011hx}, the MGS measurement from Ref.~\cite{Ross:2014qpa} and the JLA SNe Ia catalog from \cite{Betoule:2014frx}, all readily available in the \cosmomc\mbox{} code. We do not include any measurements of the Hubble constant $H_0$, apart from a uniform prior $0.4\leq h \leq 1.0$.
	
We consider an exactly flat Universe, $\OK = 0$ with a fixed He fraction $Y_{\mathrm{He}} = 0.24$ after BBN, effective number of neutrino species $N_\nu = 3.046$, with one massive neutrino $N_{\nu,m} = 0$, no sterile neutrino species $N_{\nu,s} = 0$ and sum of neutrino masses $\sum m_\nu = \unit[0.06]{eV}$. We have explicitly checked that having the neutrino flavor number and masses as free parameters does not alter our results.

The code takes as input, for every cosmological parameter going into the MCMC, a flat, bounded prior distribution centered at $x_0$ and bounded in the interval $(x_{\mathrm{min}},x_{\mathrm{max}})$, with a proposed step size $\Delta_{\mathrm{prop}}$ no greater that $\Delta_{\mathrm{max}}$.

The chosen priors for the two runs are shown on Table~\ref{tab:Priors_CosmoMC}. Priors for $\Ob h^2$, $\Oc h^2$, $\theta_{\mathrm{MC}}$, $\tau_{\mathrm{re}}$ and $\log(10^{10} A_s)$ were taken to be the default \cosmomc\mbox{} ones, while priors for $\ns$, $\as$, $\woDE$, $\waDE$, $r_{0.05}$ and $\Ninf$ were chosen with the theoretical bounds derived from \textit{Planck} satellite (2015 data release) in \cite{Ade:2015xua}. In particular,
	\begin{itemize}
	\item The approximate starting point $x_0$ was taken to be the center of the \cosmomc\mbox{} distributions $w_0w_a$CDM in Table~\ref{tab:Constraints_Cosmology}.
	\item The interval $(x_{min},x_{max})$ was taken to roughly encompass $5 \sigma$ in the \cosmomc\mbox{} $w_0w_a$CDM distributions and are shown in Table~\ref{tab:Constraints_Cosmology}.
	\item The approximate criteria chosen for the step parameters was chosen to be $\Delta_{\mathrm{max}} = |x_{\mathrm{max}} - x_{\mathrm{min}}|)/100$ and $\Delta_{\mathrm{prop}} = \Delta_{\mathrm{max}}/200$.
	\end{itemize}

The burn-in was removed by taking out the first 1500 samples in each of the chains. This is more than enough since it was checked that convergence was achieved in all cases much faster, after the 500-th sample, at most. We decided, however, to remove the extra samples to ensure that the chains left were truly Markovian.

The simulations' main features are summarized next:
	\begin{itemize}
	\item For the $w_0w_a$CDM run, we have 16 chains in total, each cut at 20k samples on average per chain (after burn-in), adding up to 320k samples in total.
	\item For the HMD run, we have 32 chains in total, each cut at 16k samples on average per chain (after burn-in), adding up to 515k samples in total.
	\end{itemize}
	
The approximate factor of two in the number of chains is due to the fact that, since the the posterior distribution of the cosmological parameters in the HDM is often non Gaussian, a larger amount of samples is required to properly capture the correlated area with \getdist.

\subsection{Constraints on cosmological parameters}
\label{subsec:Constraints_on_cosmological_Parameters}

	\begin{table*}[!t]
	\centering
	\begin{tabular} { l | r r r | r r r}
	\hline \hline
	Parameter & $w_0w_a$CDM & HDM (pred.) & HDM (obs.) & $w_0w_a$CDM & HDM (pred.) & HDM (obs.)\\
	\hline
	 & & Confidence level 68.3\% & & & Confidence level 95.5\% & \\
	\hline
	$\Omega_b h^2 $ 			& $0.02237\pm 0.00025 $ 					& $0.02231\pm 0.00022 $ 					 & $0.02233\pm 0.00022 $ 						& $0.02237^{+0.00051}_{-0.00049}$ 		& $0.02231^{+0.00043}_{-0.00043}$ 	& $0.02233^{+0.00044}_{-0.00043}$	\\
	$\Omega_c h^2 $ 			& $0.1177\pm 0.0018 $ 						& $0.1181\pm 0.0011 $ 						 & $0.1177\pm 0.0013 $ 							& $0.1177^{+0.0035}_{-0.0035}$ 			& $0.1181^{+0.0024}_{-0.0021}$ 			 & $0.1177^{+0.0025}_{-0.0025}$			\\
	$100\theta_{MC} $ 		& $1.04111\pm 0.00045 $ 					& $1.04106\pm 0.00040 $ 					 & $1.04110\pm 0.00042 $ 						& $1.04111^{+0.00088}_{-0.00088}$ 		& $1.04106^{+0.00080}_{-0.00081}$ 	& $1.04110^{+0.00083}_{-0.00082}$	\\
	$\tau_{RE}$ 					& $0.069^{+0.017}_{-0.019} $ 				& $0.066\pm 0.013 $ 							 & $0.070\pm 0.014 $ 								& $0.069^{+0.038}_{-0.035} $ 				& $0.066^{+0.025}_{-0.025} $ 				 & $0.070^{+0.027}_{-0.027} $				\\
	${\rm{ln}}(10^{10} A_s)$ 	& $3.067^{+0.032}_{-0.036} $ 				& $3.063\pm 0.025 $ 							 & $3.068\pm 0.026 $ 								& $3.067^{+0.069}_{-0.064} $ 				& $3.063^{+0.049}_{-0.049} $ 				 & $3.068^{+0.050}_{-0.050} $				\\
	$w_0 $ 							& $-0.93\pm 0.10 $ 							& $-0.99999^{+0.0025}_{-0.0020}$ 		 & $-1.0001\pm 0.0032 $ 							& $-0.93^{+0.21}_{-0.20} $ 					& $-0.99999^{+0.0056}_{-0.0060}$ 		 & $-1.0001^{+0.0072}_{-0.0074}$		\\
	$w_a $ 							& $-0.21^{+0.41}_{-0.31} $ 					& $-0.015^{+0.071}_{-0.048} $ 			 & $0.001^{+0.039}_{-0.034} $ 					& $-0.21^{+0.69}_{-0.74} $ 					& $-0.02^{+0.18}_{-0.22}$				 	 & $0.00^{+0.15}_{-0.16} $					\\
	$\ns $ 							& $0.9694\pm 0.0056 $ 						& $0.9665^{+0.0032}_{-0.0022}$ 		 & $0.9693^{+0.0046}_{-0.0042}$ 			& $0.969^{+0.011}_{-0.011} $ 					& $0.9665^{+0.0045}_{-0.0051}$ 		 & $0.9693^{+0.0083}_{-0.0082}$			\\
	$\as $ 							& $-0.0047\pm 0.0078 $ 						& $-0.0027\pm 0.0073 $ 						 & $-0.0014\pm 0.0066 $ 							& $-0.005^{+0.015}_{-0.015} $ 				& $-0.003^{+0.015}_{-0.014} $	 			 & $-0.001^{+0.013}_{-0.014} $				\\
	$r_{0.05} $ 					& $0.045^{+0.017}_{-0.038} $ 				& $0.0002\pm 0.0017 $ 						 & $0.00255^{+0.00070}_{-0.0010}$ 			& 						$< 0.0964 $ 										 & $0.0002^{+0.0031}_{-0.0033}$ 		& $0.0025^{+0.0017}_{-0.0016}$			\\
	$\Ninf $ 						& $n/a$ 											& $n/a$ 											 & $70^{+9}_{-10} $ 								&				 $n/a $ 												 & $n/a$ 											& $70^{+20}_{-20} $							\\
	\hline
	\end{tabular}
	\caption{Constraints on cosmological parameters $\Omega_b h^2$, $\Omega_c h^2$, $100\theta_{MC}$, $\tau_{RE}$, $\woDE$, $\waDE$, ${\rm{ln}}(10^{10} A_s)$, $\ns$, $\as$, $r_{0.05}$ and $\Ninf$ and \textit{BKP+len.+low$\ell$+MPK+ext.} for the $w_0w_a$CDM and HDM (observed and predicted) \cosmomc chains at the 68.3\% and 95.5\% confidence levels.}
	\label{tab:Constraints_Cosmology}
	\end{table*}
	
On a first, naive approximation, we took the $w_0w_a$CDM samples and imposed the constraints of Eqs.~\eqref{eq:1st_Order_Relation_Delta}, ~\eqref{eq:2nd_Order_Relation_Delta} and~\eqref{eq:r_Relation} directly, by selecting the samples that had values $\ns$, $\as$ and $r_{0.05}$ close to those predicted by $\woDE$, $\waDE$ and $\ODE$, that is, calculating the three spectral quantities in terms of these late Universe observables and selecting only samples within an interval
	\begin{equation}
	x_{\mathrm{sample}} \in (x_{\mathrm{cons}.} - \Delta_{x},x_{\mathrm{cons}.} + \Delta_{x})
	\;,
	\end{equation}
where $x = \ns,\as,r_{0.05}$ and $\Delta_{\ns}$, $\Delta_{\as}$ and $\Delta_{r_{0.05}}$ are three parameters set to their minimum observed value that still allows to keep a sufficiently large number of samples for posterior marginalization. The aim for this calculation is to roughly estimate what region in cosmological parameter space is allowed by the Higgs-dilaton model.

This proves to be, however, too limiting, as the HDM is extremely constraining on some of the parameters (see the figures of merit for a variety of pairs of parameters in Table~\ref{tab:FoM}), which meant, in effect, that in order to keep a number of samples greater that $\Order(10^3)$, then the tolerance parameter that controls the width of the ad-hoc constrain, $\Delta_{x}$, becomes unacceptably large.

It is necessary, then, to create a mock sampling with a much larger sample density. This was done by the following approximation: by assuming the \textit{BKP+MPK+len.+low$\ell$+ext.} likelihood is close to a Gaussian likelihood, we calculate the mean and covariance in order to then sample the resulting multinormal distribution with $\Order(10^7)$ points, which constitutes a 40-fold increase in density sampling and allows enough statistics to marginalize the subsample selected by imposing the HDM constraints on the spectral quantities.

The constraints for $\Omega_b h^2$, $\Omega_c h^2$, $100\theta_{\mathrm{MC}}$, $\tau_{\mathrm{re}}$, $\woDE$, $\waDE$, ${\rm{ln}}(10^{10} A_s)$, $\ns$, $\as$, $r_{0.05}$ and $\Ninf$ if applicable for $w_0w_a$CDM and the HDM ($20 \leq \Ninf \leq 96$) are shown on Eqs.~(\ref{eq:HDM_Couplings_Bounds_68.3}) and~(\ref{eq:HDM_Couplings_Bounds_95.5}) at the 68.3\% and 95.5\% confidence levels. Remember that the two constraints for the Higgs-dilaton model are calculated in two different ways, one naive (theoretical) and the other numerical and more robust.

The theoretically predicted HDM contours, obtained from the base $w_0w_a$CDM \cosmomc\mbox{} run along with the actual HDM contours obtained from the MCMC are shown in Figs.~\ref{fig:HDMpre+sim_Ombh2,Omch2,100theta,tau,As_} and~\ref{fig:HDMpre+sim_w0,wa,ns,as,r_}. The most interesting cosmological parameters of the models, are shown in Fig.~\ref{fig:wCDM+HDMpre+sim_w0,wa,ns,as,r,Ninf_histogram} showing the marginalized distributions for $w_0$, $w_a$, $\ns$, $\as$, $r_{0.05}$, $\Ninf$ for the base $w_0w_a$CDM cosmology, the theoretical HDM correlated regions obtained from it and the actual HDM correlated regions sampled with the MCMC. Note that the theoretically predicted plots on $r$ fail spectacularly precisely because of the non-Gaussianity of the \textit{BKP+MPK+len.+low$\ell$+ext.} likelihood on this parameter. Finally, in Fig.~\ref{fig:wCDM+HDM_w0,wa,ns,as,r_} we show the $1\sigma$ and $2\sigma$ contours for various combinations of parameters of the HDM and $w_0 w_a$CDM models, while in Fig.~\ref{fig:HDMmodified_ns,d0,as,wa_detail} we show the $1\sigma$, $2\sigma$ and $3\sigma$ contours for a selection of interesting parameters such as $n_s$ and $\alpha_s$ along with the asymptotic analytical solutions.

\subsection{Constraints on HDM couplings}
\label{subsec:Constraints_on_HDM_couplings}

\begin{table*}[!t]
\centering
\begin{tabular}{l | r r r}
\hline \hline
FoM 				& $w_0w_a$CDM 	& HDM 		& $Q_{w_0w_a\text{CDM},\text{HDM}}$ 	\\
\hline
$w_0,w_a$ 		& 8 						& 2216 		& 272 													 \\
$w_0,A_s$ 		& 40 						& 3014 		& 76 														 \\
$w_0,\ns$ 			& 240 					& 17847 	& 74 														 \\
$w_0,\as$ 			& 170 					& 12281 	& 72 														 \\
$w_0,r_{0.05}$ 	& 50 						& 99458 	& 1985 													 \\
$w_a,A_s$ 		& 13 						& 157 		& 12 														 \\
$w_a,\ns$ 			& 74 						& 940 		& 13 														 \\
$w_a,\as$ 			& 49 						& 635 		& 13 														 \\
\hline \hline
\end{tabular}
\quad
\begin{tabular}{l | r r r}
\hline \hline
FoM 						& $w_0w_a$CDM 	& HDM 		& $Q_{w_0w_a\text{CDM},\text{HDM}}$ 	\\
\hline
$w_a,r_{0.05}$ 		& 14 						& 5152 		& 358 													 \\
$A_s,\ns$ 				& 956 					& 1405 		& 1.5 													 \\
$A_s,\as$ 				& 559					& 946 		& 1.7 													 \\
$A_s,r_{0.05}$ 		& 156 					& 7334 		& 47 														 \\
$\ns,\as$ 				& 3202 					& 5424 		& 1.7 													 \\
$\ns,r_{0.05}$ 		& 951 					& 213009 	& 224													 \\
$\as,r_{0.05}$ 		& 698 					& 28838 	& 41 														 \\
 & & & \\
\hline \hline
\end{tabular}
\caption{Figures of Merit for the $w_0w_a$CDM and HDM, as well as the ratio $Q = \mathrm{FoM}_{w_0w_a\mathrm{CDM}}/\mathrm{FoM}_{\mathrm{HDM}}$ between the two quantities, giving the factor by which the HDM constraints the correlated region compared to $w_0w_a$CDM for the selected pairs of parameters.}
\label{tab:FoM}
\end{table*}
	
In this section we present the constraints for the HDM couplings $\xh/\sqrt{\lambda}$ and $\xc$ by numerically inverting Eqs.~\eqref{eq:s_Amplitude_Couplings} and~\eqref{eq:s-Index_Couplings}. Doing so, we obtain
\begin{align}
\label{eq:HDM_Couplings_Bounds_68.3}
\xc 						& < 0.00109
							\\
\xh/\sqrt{\lambda} 	& = 59200^{+6100}_{-14000}
								\;
\end{align}
at the 68.3\% confidence level, and
	\begin{align}
	\label{eq:HDM_Couplings_Bounds_95.5}
	\xc 						& < 0.00328
								\\
	\xh/\sqrt{\lambda} 	& = 59200^{+30000}_{-20000}
								\;
	\end{align}
at the 95.5\% and confidence level.

\begin{figure*}[!t]
\centering
\includegraphics[width = .7\textwidth]{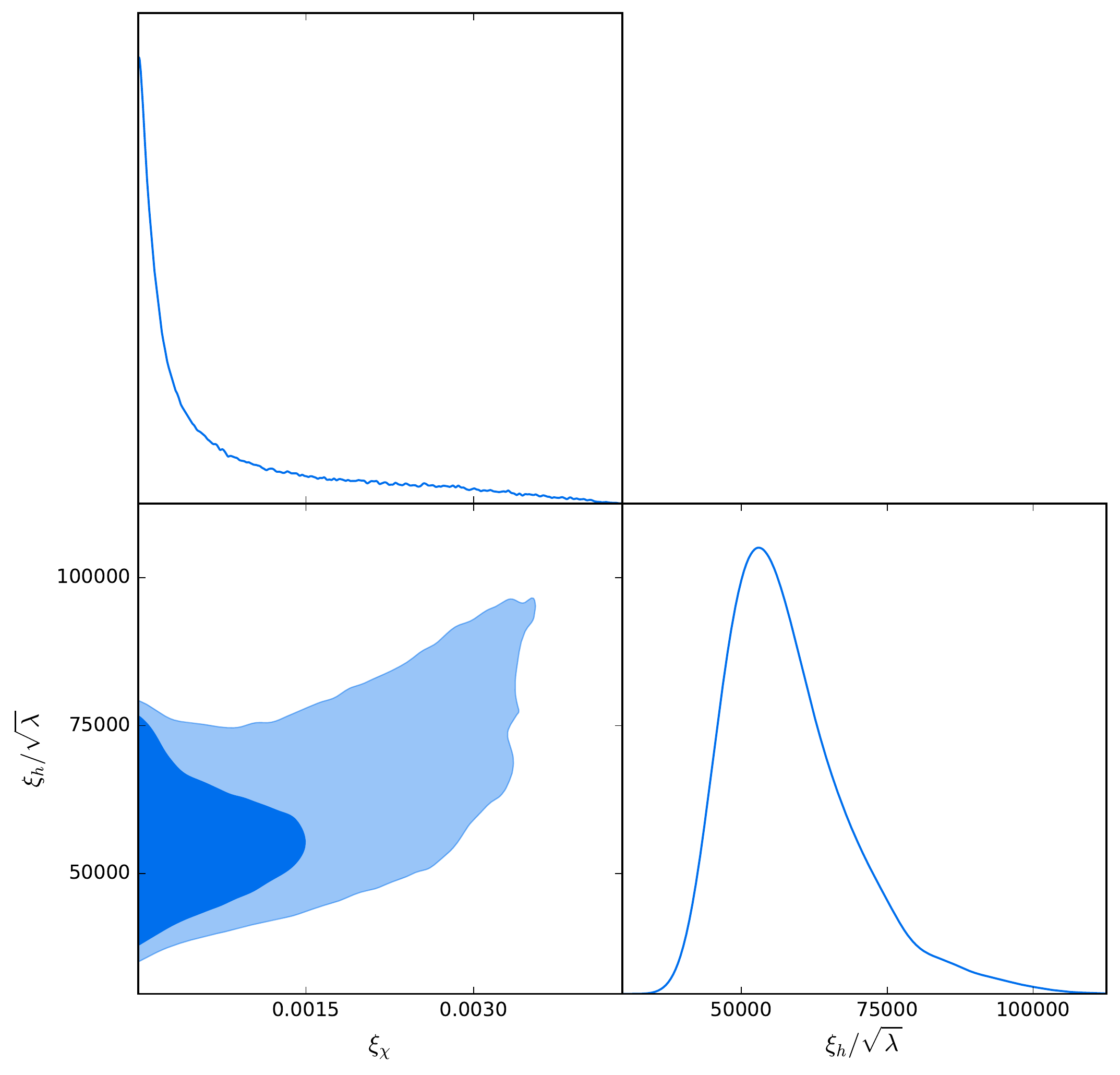}
\caption{$1\sigma$ and $2\sigma$ contours of the HDM couplings $\xh$, $\xc$ and $\lambda$ from the \textit{BKP+len.+low$\ell$+MPK+ext.} obtained from the HDM \cosmomc chains at the 68.3\%, 95.5\% and 99\% confidence levels.}
\label{fig:HDM_couplings}
\end{figure*}

Finally, we present the $68.3\%$ and $95.5\%$ confidence contours for $\xc$ and $\xh/\sqrt{\lambda}$ on Fig.~\ref{fig:HDM_couplings}. As it can be seen, there is significant improvement over the predictions of the confidence intervals with respect to Ref.~\cite{GarciaBellido:2011de}.

\subsection{Bayesian analysis of $w_0w_a$CDM vs. HDM}
\label{subsec:Bayesian_analysis_of_w0waCDM_vs_HDM}

We proceed now to compare the two models, $w_0w_a$CDM vs. Higgs-dilaton Inflation, in detail. It should be first noted that, in Table~\ref{tab:Constraints_Cosmology} and Fig.~\ref{fig:wCDM+HDMpre+sim_w0,wa,ns,as,r,Ninf_histogram}, while the HDM makes very specific and tight constraints on a number of parameters, all of these intervals are completely within the $1\sigma$ values of $w_0w_a$CDM and thus, the two models are as of today indiscernible one from another, though future measurements by \textit{Euclid}, \textit{DES} and \textit{PAU} may be able to change that.

We find that the HDM is on an equal footing with respect to the $w_0w_a$CDM model as they have a similar chi-square, i.e. $\Delta \chi^2 = \chi^2_{w_0w_a\mathrm{CDM}} - \chi^2_{\mathrm{HDM}} = 0.178$, but with the HDM model having two fewer parameters, thus being equally compatible with the $\Lambda$CDM model, nested itself in $w_0w_a$CDM.

Furthermore, we also estimate the Bayesian evidence of the two models and compare them by making use of the Jeffreys' scale. The Bayesian evidence $E(\mathbf{D} \vert \mathcal{M})$ for parameters $u$ of a model $\mathcal{M}$ given some data $\mathbf{D}$ is defined as an integral of the likelihood distribution function over the parameter priors, e.g. see Ref.~\cite{Nesseris:2012cq}, that is
\begin{equation}
\label{eq:Bayes_Evidence}
E(\mathbf{D} \vert \mathcal{M}) = \int_\mathcal{R} du~\mathcal{L} (\mathbf{D} \vert u,\mathcal{M}) \pi (u,\mathcal{M})
\;,
\end{equation}
where $\mathcal{L} (\mathbf{D} \vert u,\mathcal{M}) = \exp(-\chi^2(u)/2)$ is the likelihood distribution function within a given $\pi(u,\mathcal{M})$ set of priors  bounding the region $\mathcal{R}$ in parameter space.
	
Unfortunately, an exact computation of the Bayesian evidence by numerically integrating the MCMC chains obtained by a Monte Carlo method was in our case numerically sensitive to the choice of certain parameters controlling for the precision of the integral in Eq.~\eqref{eq:Bayes_Evidence}. This is due to the fact that the evidence is very sensitive to the tails of the likelihood distribution function. A precise computation of the Bayesian evidence requires the MCMC sampling of the full prior region in parameter space to accurately sample the tails, but the typical MCMC sampler, and in our case \cosmomc\mbox{}, only samples the posterior correlated region, thus leaving the tails barely sampled. Clearly, this leads to the aforementioned numerical sensitivity of the Bayesian evidence computation on the integration method.

Fortunately, there is a number of ways to circumvent this issue and here we compute three of them:
\begin{enumerate}
  \item A well-established approximation to the Bayesian evidence by using the Harmonic Mean Approximation (HMA) for the evidence ratio, see Ref.~\cite{Weinberg:2009rd}.
  \item The Akaike Information Criterion (AIC), see Ref.~\cite{Liddle:2007fy}.
  \item The Deviance Information Criteria (DIC), see Ref.~\cite{Liddle:2007fy}.
\end{enumerate}
These are respectively defined as follows:
\begin{align}
\label{eq:Bayes_Evidence_HMA}
E^{\mathrm{HMA}}(\mathbf{D} \vert \mathcal{M}) =	& \left( \frac{1}{N} \sum_{i = 1}^N \frac{1}{\mathcal{L} (\mathbf{D} \vert u,\mathcal{M})} \right)^{-1}
																				\\
\label{eq:Bayes_Evidence_AIC}
\mathrm{AIC}(\mathbf{D} \vert \mathcal{M}) =			& 2 k +\chi^2_{\mathrm{min}} (\mathbf{D} \vert u,\mathcal{M})
																			\\													 
\label{eq:Bayes_Evidence_DIC}
\begin{split}
\mathrm{DIC}(\mathbf{D} \vert \mathcal{M}) =& 2 \langle \chi^2 (\mathbf{D} \vert u,\mathcal{M})\rangle - \chi^2_{\mathrm{min}} (\mathbf{D} \vert u,\mathcal{M})
\end{split}
\end{align}
where $N$ in the number of samples in the MCMC chains, $k$ the number of parameters to be estimated and the $\langle \cdots \rangle$ denote an average over the posterior distribution.

In order to compare two models $\mathcal{M}_1$ and $\mathcal{M}_2$, we can make use of the ratio of the Bayesian evidences calculated by the Harmonic Mean Approximation (see Ref.~\cite{Nesseris:2012cq}), given by:
\begin{equation}
\label{eq:Bayes_Evidence_Ratio}
R^{\mathrm{HMA}}_{1,2} = \frac{E^{\mathrm{HMA}}(\mathbf{D} \vert \mathcal{M}_1)}{E^{\mathrm{HMA}}(\mathbf{D} \vert \mathcal{M}_2)}
\;,
\end{equation}
or alternatively, their differences in the AIC or DIC criteria
\begin{align}
\label{eq:Bayes_Evidence_Difference_AIC,DIC}
\Delta \mathrm{AIC}_{1,2} 	& = \mathrm{AIC}(\mathbf{D} \vert \mathcal{M}_1) - \mathrm{AIC}(\mathbf{D} \vert \mathcal{M}_2)
											\\
\Delta \mathrm{DIC}_{1,2} 	& = \mathrm{DIC}(\mathbf{D} \vert \mathcal{M}_1) - \mathrm{DIC}(\mathbf{D} \vert \mathcal{M}_2)
											\;.
\end{align}
	
Making use of the approximations of Eqs.~\eqref{eq:Bayes_Evidence_HMA}, ~\eqref{eq:Bayes_Evidence_AIC}, ~\eqref{eq:Bayes_Evidence_DIC} and the definition of Eq.~\eqref{eq:Bayes_Evidence_Ratio}, we find the Bayesian evidence ratios and AIC/BIC differences for HDM and $w_0w_a$CDM to be
\begin{align}
R^{\mathrm{HMA}}_{\mathrm{HDM},w_0w_a\mathrm{CDM}} = 		&~~0.55 \approx 1.8^{-1}
																									\\
\Delta \mathrm{AIC}_{\mathrm{HDM},w_0w_a\mathrm{CDM}} = 	& - 4.2,
																									\\
\Delta \mathrm{DIC}_{\mathrm{HDM},w_0w_a\mathrm{CDM}} = 	&- 3.5
																									\;,
	\end{align}
which imply that the Higgs-dilaton and the evolving dark energy models are more or less on an equal footing as seen by the evidence ratio $R^{\mathrm{HMA}}_{\mathrm{HDM},w_0w_a\mathrm{CDM}}$, but as seen by Jeffrey's scale, e.g. see Ref.~\cite{Nesseris:2012cq}, there is some evidence against the latter model. We stress, however that the evidence ratio is based on an approximate calculation and as such should be interpreted with caution.

Finally, in order to quantify by which amount the parameter space is constrained, we use the figure of merit (FoM) metric, see Ref. \cite{Albrecht:2006um,Nesseris:2012cq} for more details. Note, that we define the FoM here as the inverse of the area inside the $1\sigma$ contour. Then, we define the quotient $Q(\mathrm{FoM}_i,\mathrm{FoM}_j)$ as
\begin{equation}
\label{eq:FoM}
Q \equiv \frac{\mathrm{FoM}_{\mathrm{HDM}}}{\mathrm{FoM}_{w_0w_a\mathrm{CDM}}} \equiv \frac{A_{w_0w_a\mathrm{CDM}}}{A_{\mathrm{HDM}}}
\;,
\end{equation}
so the greater the FoM, the more constrained the model is, and the value of $Q(\mathrm{FoM}_i,\mathrm{FoM}_j)$, indicates the n-fold improvement in constraining the correlated regions, that is, the quotient of the areas. In Table~\ref{tab:FoM} we give the FoM for all pairs of parameters $w_0$, $w_a$, $A_s$, $\ns$, $\as$ and $r_{0.05}$. Note that the FoM for $(w_0,w_a)$ and $(\ns,r_{0.05})$ is $\Order(10^2)$, while that of $(w_0,r_{0.05})$ is of order $\Order(10^3)$. Clearly, this illustrates why were the vast majority of samples in $w_0w_aCDM$ rejected when we naively imposed all three constraints directly on the $w_0w_a$CMD \cosmomc\mbox{} run.
\begin{figure*}[!t]
\centering
\includegraphics[width = 0.9\textwidth]{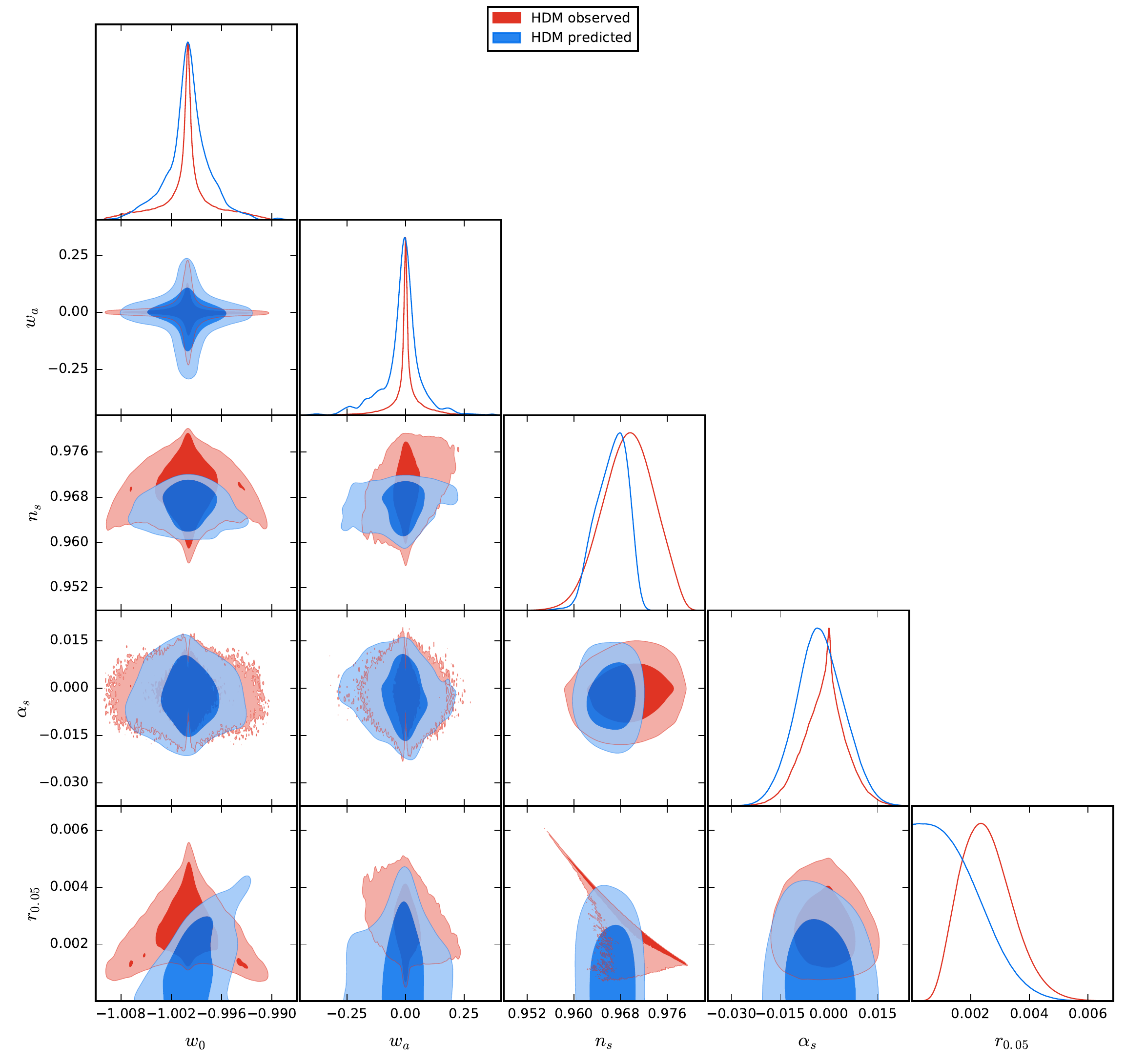}
\caption{1$\sigma$ and $2\sigma$ contours of $\Ob h^2$, $\Oc h^2$, $100\theta_{MC}$, $\tau_{RE}$ and $A_s$ for the HDM both predicted (blue) and observed (red).}
\label{fig:HDMpre+sim_Ombh2,Omch2,100theta,tau,As_}
\end{figure*}

\begin{figure*}[!t]
\centering
\includegraphics[width = 0.9\textwidth]{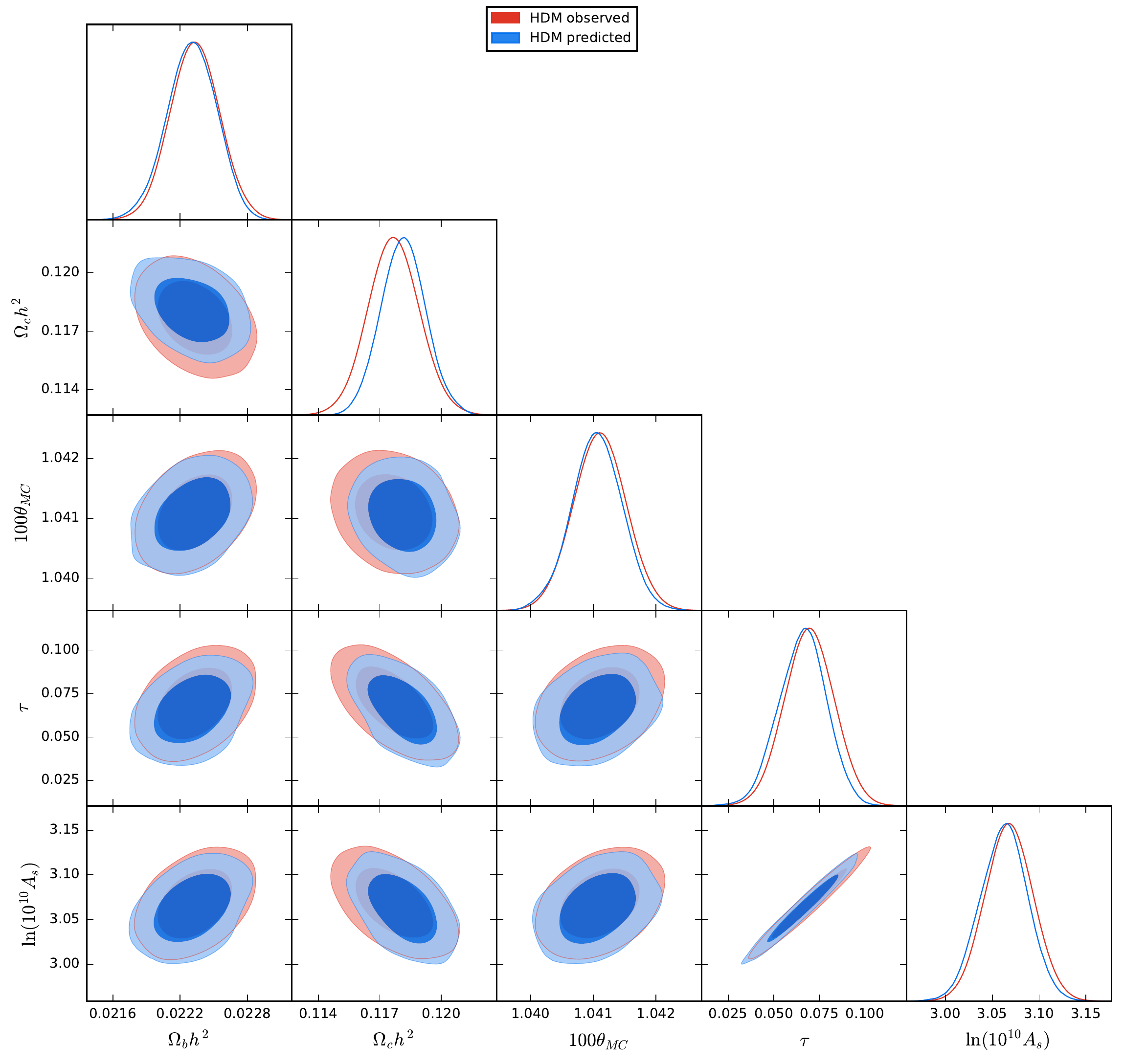}
\caption{1$\sigma$ and $2\sigma$ contours of $\woDE$, $\waDE$, $\ns$, $\as$, and $r_{0.05}$ for the HDM both predicted (blue) and observed (red). Note that the predictions on the contours involving the tensor-to-scalar ratio $r_{0.05}$, since that the assumption of approximately Gaussian clearly likelihood breaks down.}
\label{fig:HDMpre+sim_w0,wa,ns,as,r_}
\end{figure*}

\begin{figure*}[!t]
\centering
\includegraphics[width = 0.9\textwidth]{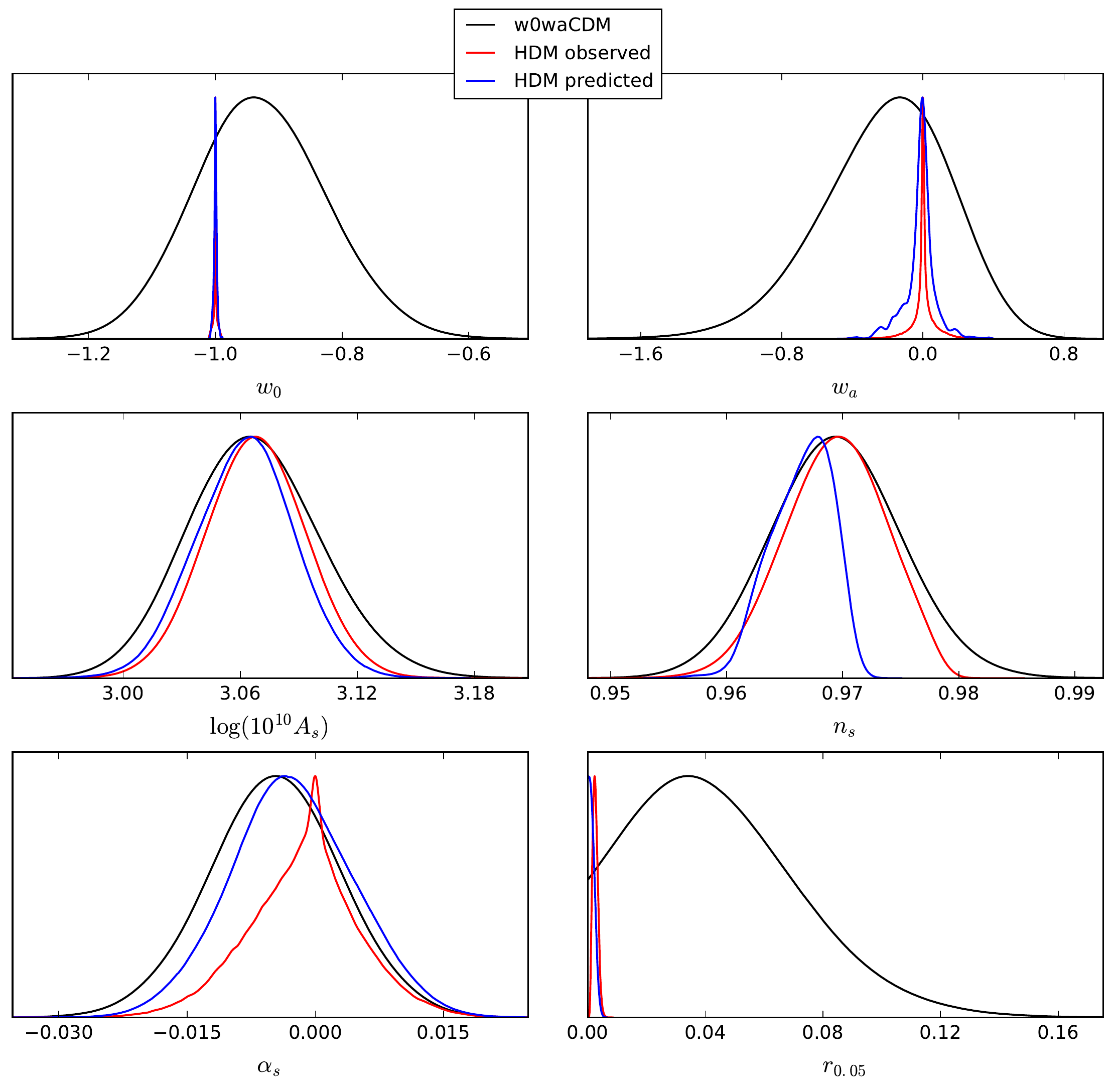}
\caption{1$\sigma$ and $2\sigma$ histograms of $\woDE$, $\waDE$, $\ns$, $\as$, and $r_{0.05}$ for the $w_0w_a$CDM model (black) and HDM both predicted (blue) and observed (red).}
\label{fig:wCDM+HDMpre+sim_w0,wa,ns,as,r,Ninf_histogram}
\end{figure*}

\begin{figure*}[!t]
\centering
\includegraphics[width = 0.9\textwidth]{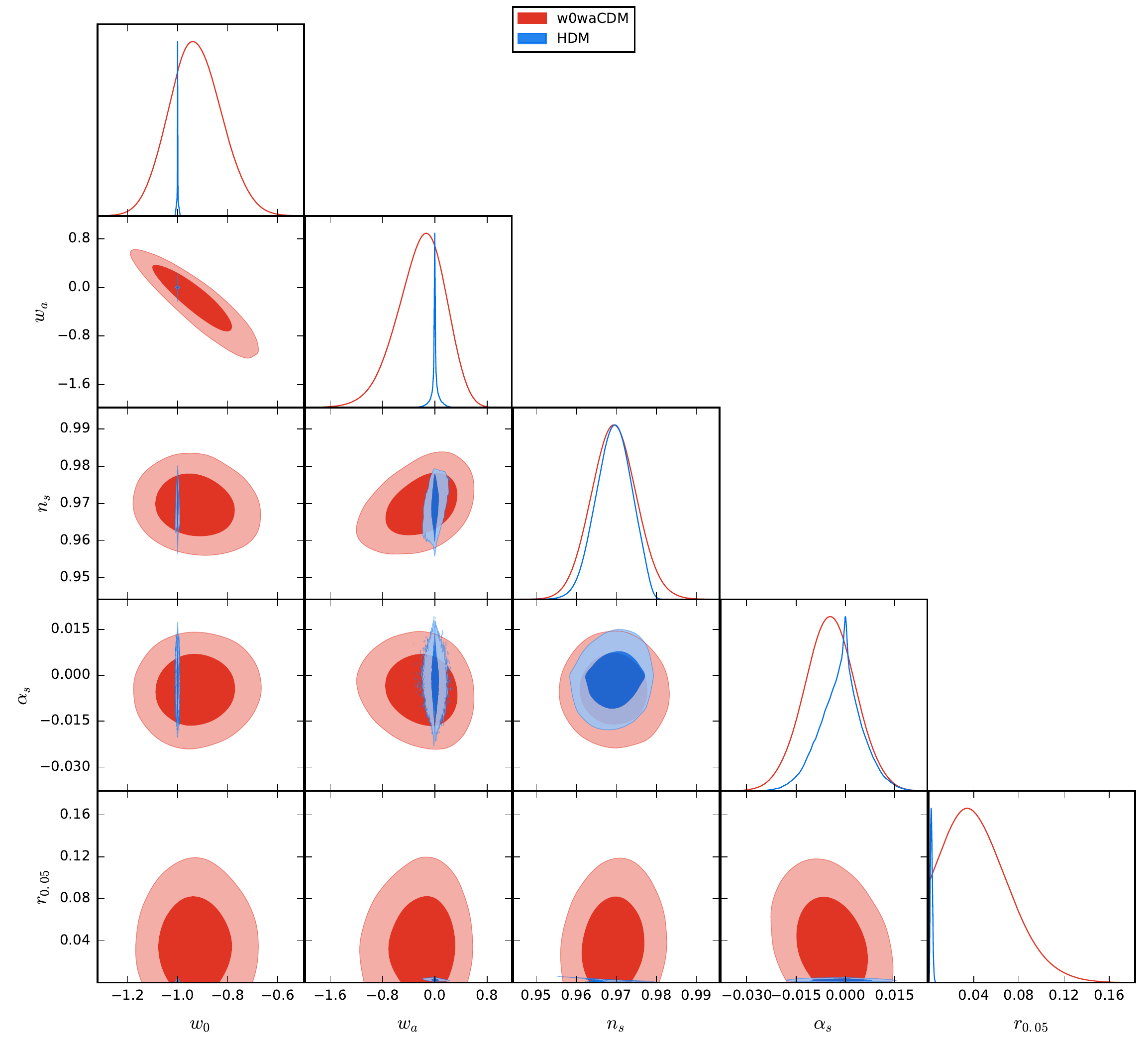}
\caption{1$\sigma$ and $2\sigma$ contour plots for $\woDE$, $\waDE$, $\ns$, $\as$, and $r_{0.05}$ for the HDM (blue) and $w_0w_a$CDM model (red), showing the great extent to which the HDM constrains the availiable cosmological parameter space for the tensor-to-scalar ratio $r_{0.05}$ as well as for the dark energy equation of state parameters  $\woDE$ and $\waDE$.}
\label{fig:wCDM+HDM_w0,wa,ns,as,r_}
\end{figure*}

\begin{figure*}[!t]
\centering
\includegraphics[width = 0.44\textwidth]{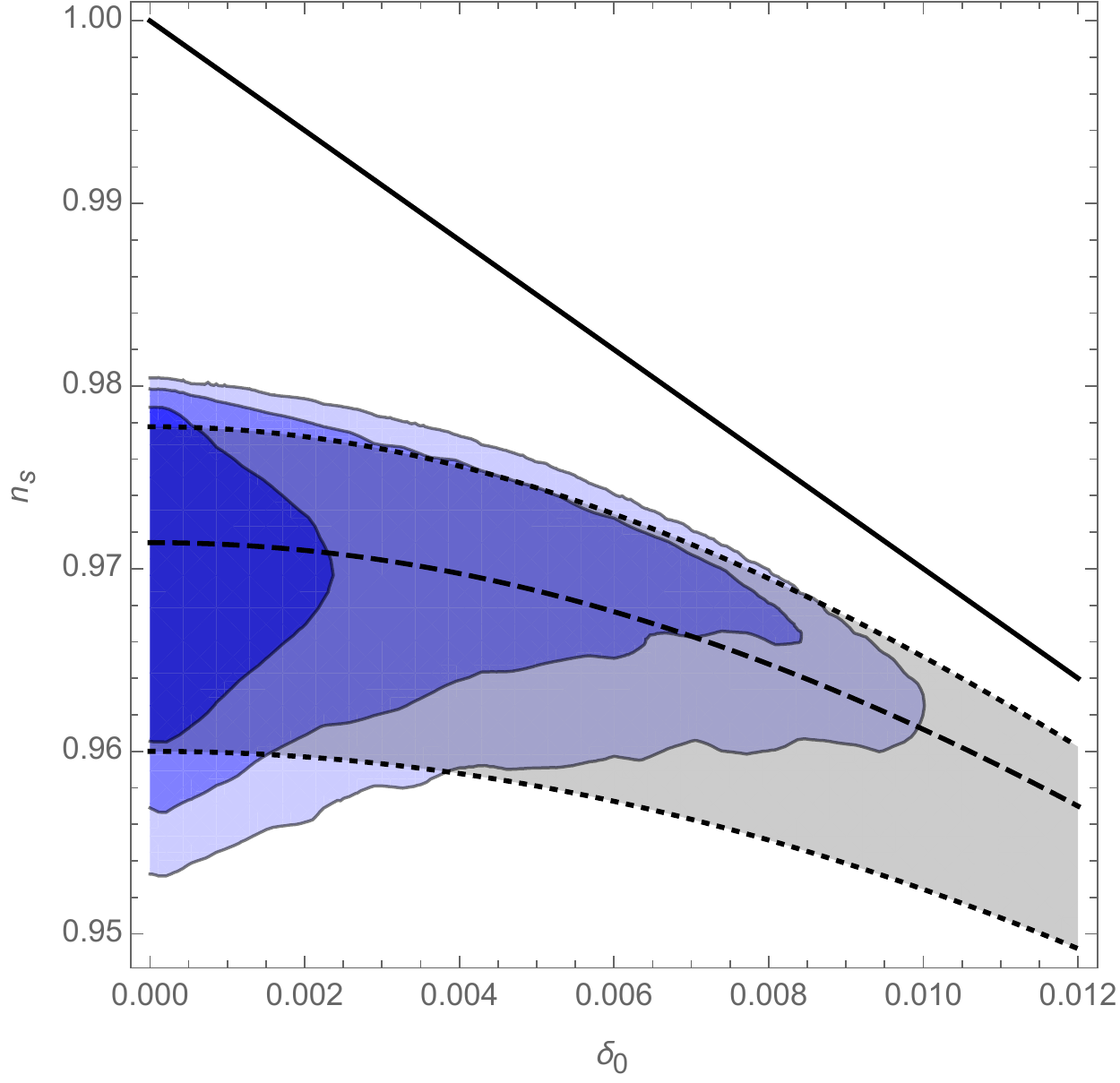}
\includegraphics[width = 0.44\textwidth]{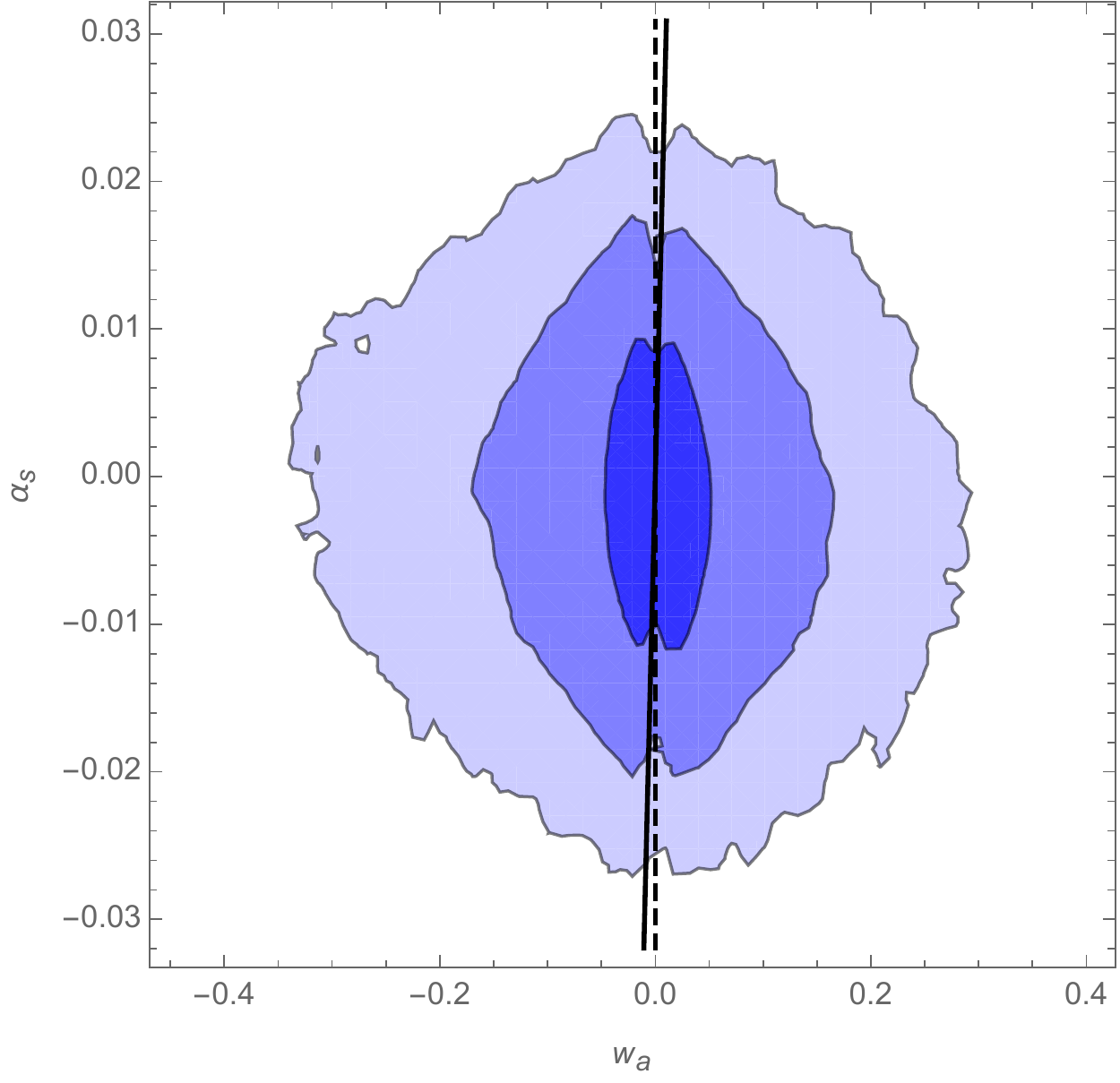}
\caption{1$\sigma$, $2\sigma$ and $3\sigma$ contours of $\dDE$, $\ns$ (left) and $\waDE$, $a_s$ (right) for the HDM. The solid line corresponds to the asymptotic solutions $\ns = 1 - 3 \dDE$ (left) and $\as = 3 \waDE$ (right). In both plots, the dashed line corresponds to the expressions of $\ns$ and $\as$ in Eqs.~\eqref{eq:1st_Order_Relation_Delta} and~\eqref{eq:2nd_Order_Relation_Delta} for the case of $\Ninf = 70$ while the dotted lines corresponds to the expressions of $\ns$ and $\as$ in Eqs.~\eqref{eq:1st_Order_Relation_Delta} and~\eqref{eq:2nd_Order_Relation_Delta} for the $2 \sigma$ bounds on the number of e-folds, that is $\Ninf = 50$ (lower bound) and $\Ninf = 90$ (upper bound) respectively.}
\label{fig:HDMmodified_ns,d0,as,wa_detail}
\end{figure*}

\section{Conclusions}
\label{sec:Conclusions}

In this paper we have found that the Higgs-dilaton model (HDM) is a viable extension of the Standard Model, one that based only on scale invariance and a nonminimal coupling to Unimodular Gravity is able to produce an early-Universe period of inflation eventually as well as explaining the present era of accelerated expansion. We have compared this model to the cosmological constant ($\Lambda$CDM) and evolving dark energy ($w_0 w_a$CDM) models, by using the latest cosmological data that includes the cosmic microwave background temperature, polarization and lensing data from the \textit{Planck} satellite, the BICEP and Keck Array experiments, the Type Ia supernovae from the JLA catalog, the Baryon Acoustic Oscillations and finally, the Weak Lensing data from the CFHTLenS survey, by implementing the model constrains in \cosmomc, a MCMC code.

It should be stressed that the links between the observables $\ns$ and $\as$, related to inflation, and $\woDE$ and $\waDE$, related to dark-energy, are very particular predictions of the model that relate two seemingly totally independent epochs, establishing a measurable relation between observables from CMB anisotropies with a largely unknown dark-energy sector.

Also, one possible issue for the HDM is that given the large values of the Higgs self-coupling of Eqs.~(\ref{eq:HDM_Couplings_Bounds_68.3}), and the wide temperature interval of Eq.~(\ref{eq:Reheating_Temperature_Bounds}), the model could in principle suffer from the strong coupling  \cite{Bezrukov:2010jz}, which could possibly affect the early time cosmology \cite{Bezrukov:2011sz}. However, the results we quoted in the aforementioned equations have always been expressed in terms of the effective $\xi_h/\sqrt\lambda$ during inflation, and therefore this statement depends on the Renormalization Group Equations (RGE) running of both $\xi_h$ and $\lambda$. In certain models $\lambda$ can be sufficiently fine-tuned to be small enough that the Higgs non-minimal coupling is not strongly coupled, and therefore all the vacuum stability and consistency of the model is not questionable.

Regarding the vacuum stability we should note that the Higgs inflation is somewhat different in that for example, in $R^2$-inflation the $R^2$ term effectively introduces another scalar field and this will help the SM stability bounds. However, as it was shown in Ref.~\cite{Bezrukov:2014ipa}, the Higgs inflation model can successfully work even if the SM vacuum is not absolutely stable, something which is another advantage of the model.

We should note that even if the RGE running of $\lambda$ and $\xi_h$, see Ref. \cite{Herranen:2014cua}, may shift these values to accommodate $\sqrt\lambda/\xi_h \sim 10^-5$, it is reasonable to assume that $\lambda$ does not reach $10^{-13}$ without some fine-tuning, so we do expect somewhat large values of $\xi_h \sim 100$. In this case the scale of new strong gravity corrections becomes dangerously close to the inflationary scale, and a proper treatment of UV corrections is needed, see Refs. \cite{Bezrukov:2011sz,Bezrukov:2012hx}.

Furthermore, the RGE equations determine the running of all SM couplings, and in particular the Higgs self-coupling $\lambda$ and its non-minimal coupling $\xi_h$, as well as the top quark mass. This last quantity is unknown to sufficient accuracy from LHC data to determine the sign of $\lambda$ at $10^{15}$ GeV. It is perfectly possible that it remains positive and, moreover, it could be small enough to alleviate the issues discussed above. This issue has been fully explored in Ref.~\cite{Herranen:2014cua}. In fact, the presence of extra scalars (although the dilation is a special one, due to its derivative coupling to the Higgs) tend to favor the stability of the SM vacuum.

We found that the values of all cosmological parameters allowed by the Higgs-dilaton model Inflation are well within the $w_0 w_a$CDM constraints. In particular, we found that $\woDE = -1.0001^{+0.0072}_{-0.0074}$, $\waDE = 0.00^{+0.15}_{-0.16}$, $\ns = 0.9693^{+0.0083}_{-0.0082}$, $\as = -0.001^{+0.013}_{-0.014}$ and $r_{0.05} = 0.0025^{+0.0017}_{-0.0016}$ (95.5\%C.L.). We also placed new stringent constraints on the couplings of the Higgs-dilaton model to gravity and  found that $\xc < 0.00328$ and $\xh/\sqrt{\lambda} = 59200^{+30000}_{-20000}$ (95.5\%C.L.). All of these are very relevant predictions of the model that may soon be confronted with observational data by \textit{DES}, \textit{PAU} and \textit{Euclid}.

Furthermore, we found that the HDM is at a slightly better footing than the $w_0 w_a$CDM model, as they both have practically the same \textit{chi-square}, i.e. $\Delta \chi^2 = \chi^2_{w_0 w_a\mathrm{CDM}}-\chi^2_{\mathrm{HDM}}=0.18$, but with the HDM model having two parameters less, and finally a Bayesian evidence favoring equally the two models but the HDM being preferred by the AIC and DIC information criteria. All of these constitute a significant improvement with respect to the analysis previously carried out in Ref.~\cite{GarciaBellido:2011de}.


\begin{acknowledgments}
We would like to thank Andres D\'{\i}az-Gil, Jason Dosset, Anthony Lewis, Chia-Hsun Chuang and Claudia Scoccola for fruitful discussions, Misha Shaposhnikov and Andrei Linde for continuing encouragement with this project and an anonymous referee for very constructive comments. We also acknowledge use of the Hydra cluster at the Instituto de F\'{\i}sica Te\'{o}rica (IFT), on which the numerical computations for this paper took place.

This work is supported by the Research Project of the Spanish MINECO, FPA2013-47986-03-3P, and the Centro de Excelencia Severo Ochoa Program SEV-2012-0249. S.~N. is supported by the Ram\'{o}n y Cajal programme through Grant No. RYC-2014-15843.
\end{acknowledgments}

\clearpage

\bibliographystyle{utcaps}
\bibliography{main}

\end{document}